\documentclass[3p,times]{elsarticle}

\makeatletter
\def\ps@pprintTitle{%
  \let\@oddhead\@empty
  \let\@evenhead\@empty
  \let\@oddfoot\@empty
  \let\@evenfoot\@oddfoot
}
\makeatother

\usepackage{amssymb}
\usepackage{amsmath}
\usepackage{booktabs}   % tables
\usepackage{xcolor}
\usepackage{algorithmicx}
\usepackage{algpseudocode}
\usepackage[detect-all]{siunitx}
\usepackage{soul}

\newcommand{\ie}{i.e.}

\newcommand{\Fig}{Fig.~}
\newcommand{\Figs}{Figs.~}
\newcommand{\tab}{Table~}
\newcommand{\mov}{Video~}
\newcommand{\sect}{Section~}

\usepackage{enumitem}
\newlist{inlinenum}{enumerate*}{1}
\setlist*[inlinenum,1]{%
label={\roman*},
afterlabel={)~}
}

\usepackage{amsbsy}
\newcommand{\vct}[1]{\boldsymbol{#1}}           % Vector
\newcommand{\dct}[1]{\mathbf{#1}}               % Director
\newcommand{\eb}[1]{\dct{e}_{#1}}

\newcommand{\diff}{\,\mathrm{d}}                  % Differential d for integration variable
\usepackage[bb=stix]{mathalpha}                 % mathbb low case
\newcommand{\ee}{\mathbb{e}}
\newcommand{\ii}{\mathbb{i}}

\newcommand{\fit}{^{\textsf{fit}}}
\newcommand{\crit}{^{\textrm{cr}}}

\usepackage{amsthm}

\makeatletter
\newcommand\setcurrentname[1]{\def\@currentlabelname{#1}}
\makeatother

\usepackage{caption}
\newlist{captionlist}{enumerate*}{1}
\setlist*[captionlist,1]{%
label={\alph*},
afterlabel={)~},
font=\bfseries\sffamily
}

\usepackage{totcount}
\newtotcounter{movie}
\renewcommand{\themovie}{S\arabic{movie}}

\newenvironment{movie}[1][]{
    \refstepcounter{movie}
    \noindent
    \textbf{Video \themovie. #1} \hfil\break
}{\bigskip}

\usepackage{hyperref}

\begin{document}

\begin{frontmatter}

\title{Minimal actuation and control of a soft hydrogel swimmer from flutter instability}

\author[aff1]{Ariel Surya Boiardi}
\ead{aboiardi@sissa.it}

\author[aff1]{Giovanni Noselli\texorpdfstring{\corref{cor1}}{}}
\ead{gnoselli@sissa.it}
\cortext[cor1]{Corresponding author}

\address[aff1]{SISSA\,--\,International School for Advanced Studies, Mathematics Area, via Bonomea 265, 34136 Trieste, Italy}
    
\begin{abstract}
    Micro-organisms propel themselves in viscous environments by the periodic, nonreciprocal beating of slender appendages known as flagella. 
    Active materials have been widely exploited to mimic this form of locomotion. However, the realization of such coordinated beating in biomimetic flagella requires complex actuation modulated in space and time. 
    We prove through experiments on polyelectrolyte hydrogel samples that directed undulatory locomotion of a soft robotic swimmer can be achieved by untethered actuation from a uniform and static electric field. 
    A minimal mathematical model is sufficient to reproduce, and thus explain, the observed behavior. 
    The periodic beating of the swimming hydrogel robot emerges from flutter instability thanks to the interplay between its active and passive reconfigurations in the viscous environment. 
    Interestingly, the flutter-driven soft robot exhibits a form of electrotaxis whereby its swimming trajectory can be controlled by simply reorienting the electric field. 
    Our findings trace the route for the embodiment of mechanical intelligence in soft robotic systems by the exploitation of flutter instability to achieve complex functional responses to simple stimuli. 
    While the experimental study is conducted on millimeter-scale hydrogel swimmers, the design principle we introduce requires simple geometry and is hence amenable for miniaturization via micro-fabrication techniques.
    We believe it may also be transferred to a wider class of soft active materials.
\end{abstract}  


\newpageafter{abstract}
    
\end{frontmatter}

\section{Introduction}
\label{INTRO}
At the micro-scale, single-celled organisms propel themselves in viscous environments by the motion of cilia and flagella. These exhibit periodic beating that is powered by the conversion of chemical energy into mechanical work taking place in motor proteins~\cite{oiwa_2003, nicastro_2014, nicastro_2018}.
As dictated by the constraints of low Reynolds number hydrodynamics, their motion is nonreciprocal to allow for the motility of lower-level organisms, transport phenomena, and feeding flows~\cite{purcell_1977,Lauga2011_Life,Yuan2021}. 

Artificial active materials have been widely employed to reproduce flagellar beat in bio-inspired artificial micro-swimmers~\cite{Dreyfus2005_Microscopic,Lum2016_Shape,Kaynak2017_Acoustic} and ciliary arrays in lab-on-a-chip devices~\cite{Toonder2008_Artificial,Shields2010_Biomimetic,Vilfan2010_Self,Dong2020_Bioinspired}. 
An important limitation of these solutions comes from the fact that the realization of periodic, nonreciprocal motion typically requires complex actuation by external stimuli modulated in space and time. 

Progress in the understanding of the functioning of cilia and flagella goes side by side with the development of technological solutions to mimic them in artificial structures.
In this regard, mathematical models have elucidated the role in flagellar beating of intrinsic feedback mechanisms regulating the activity of motor proteins~\cite{lesich_2010, bayly_2015,howard_2016}.
In parallel, design principles have emerged in robotics, such that the actuation of soft machines by simple control can be achieved via structural and material design.
For instance, materials undergoing cyclic chemical reactions have successfully been employed to build fully autonomous soft-robots~\cite{Maeda2007_SelfWalking}. 
Exploitation of feedback loops based on self-shadowing was recently proposed as a strategy to generate self-sustained oscillations in photo-sensitive  structures~\cite{zhao_2019,korner_2020}. Another relevant example comes from artificial swimmers consisting of a spherical particle endowed with passive flagella and driven by Quincke rotation through an electrohydrodynamic instability~\cite{stone_2020,stone_2021}.

\begin{figure}[t]
    \centering
    \includegraphics[width=\linewidth]{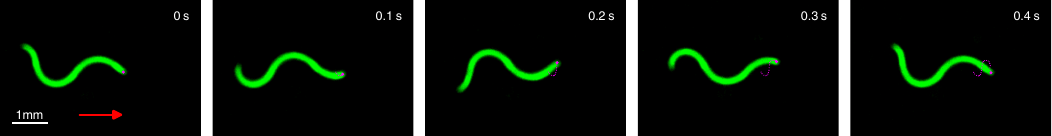}
    \caption{A polyelectrolyte hydrogel ribbon swimming through self-sustained, nonreciprocal oscillations in a solution of sodium chloride in water when subjected to a constant and uniform electric field $\vct{E}$ of sufficient magnitude (red arrow). The snapshots depict a cycle of shape change corresponding to \qty{0.4}{\second}.}
    \label{fig:intro}
\end{figure}

In this study, we propose a new mechanism for the undulatory locomotion of artificial swimmers requiring a simple environmental stimulus.
Specifically, we report  millimeter-scale physical experiments to prove that uniform ribbons of polyelectrolyte hydrogel (PEH) can swim by nonreciprocal bending oscillations when subject to a uniform and static electric field of sufficient magnitude, see \Fig\ref{fig:intro} and \mov\ref{MOV:exp_swimming}.
The wave-like motion of the soft robot closely resembles that of eukaryotic flagella in micro-organisms. Also, its performance is remarkable with average swimming speed up to \qty{1.8}{\mm\per\second} (corresponding to $\approx 0.5$ body length per second) while oscillating at frequencies between \qty{2}{\Hz} and \qty{3}{\Hz}.

Regarding the active bending of the PEH ribbon, this is caused by differential swelling along its cross-section as due to the transport of solvated ionic species driven by the electric field~\cite{Doi1992,Hong2010}; it is thus dependent upon the relative orientation of the ribbon to the external field. 
We combine resistive force theory (RFT)~\cite{Gray1955,Lighthill1976,koens_2017} with rod's morphoelasticity~\cite{Goriely2017} to incorporate such active response in a mathematical model for the planar motion of PEH ribbons in a viscous fluid. 
In light of this minimal model, we interpret the periodic oscillations observed in the experiments as limit cycles originating from flutter instability~\cite{Ziegler1977,bigoni2023flutter}.

Our analysis reveals that the key factor underlying the observed phenomenon is the subtle interplay between the active and the passive reconfigurations of the PEH ribbon, the latter arising from its elasticity and hydrodynamic interactions with the fluid environment. 
The ability of the active ribbons to generate a complex response to a minimal stimulus is thus an instance of embodiment of mechanical intelligence, whereby harnessing flutter instability allows to offload low-level control tasks into the interactions of the swimmer with the environment. 
The proposed model accurately captures all the essential features of the behavior observed in the physical experiments. 
Remarkably, by exploiting such a model to further investigate the dynamics of the robotic swimmer, we show that it can be steered in the fluid domain by simply reorienting the external field.

\section{Fabrication of PEH samples and swimming experiments under uniform and constant electric field}
\label{EXP}

Millimeter-scale, polyelectrolyte hydrogel samples with a ribbon geometry were synthesized in poly-acrylamide-\emph{co}-sodium acrylate (PAAm-\emph{co}-SA) adapting a UV photo-lithographic technique from~\cite{Damioli2022}.
The pre-gel mixture was prepared from stock solutions as reported in \tab\ref{MANDM:tab:stock_solutions_reagents}.
In particular, acrylamide and bis-acrylamide were used as monomer and cross-linker, respectively, whereas sodium acrylate was used to render the hydrogel responsive to the electric field, by introducing fixed ions in the polymeric network. 
The polymerization and cross-linking reactions were initiated by the VA-086 photoinitiator. 
Fluorescent microspheres of \qty{0.1}{\micro\m} in diameter were added as a dye, allowing to record high contrast images from the experiments.

\begin{table*}[b]
    \centering
    \caption{Reagents for the preparation of \qty{1}{\mL} of pre-gel solution. 
    Solutions are as purchased or prepared from powders using DI water (type I, ASTM D1193-91).
    }
    \label{MANDM:tab:stock_solutions_reagents}
    \begin{tabular}{lr}
        Stock solutions & Quantity\\
        \midrule
        Acrylamide (40\% w/v), Fisher Bioreagents n.\,BP1402-1          & \qty{276}{\ul} \\
        Bis-acrylamide (2\% w/v), Fisher Bioreagents n.\,BP1404-250     & \qty{254}{\ul} \\
        Sodium acrylate (10\% w/v), Sigma-Aldrich n.\,408220-25G        & \qty{230}{\ul} \\
        VA-086 (3\% w/v), Wako Chemicals n.\,013-19342                  & \qty{157}{\ul} \\
        Fluorescent microspheres (2\% w/v), Invitrogen n.\,F8803        & \qty{83}{\ul} \\
        \bottomrule
    \end{tabular}
\end{table*}

\begin{figure*}[t]
    \centering
    \includegraphics[width=\linewidth]{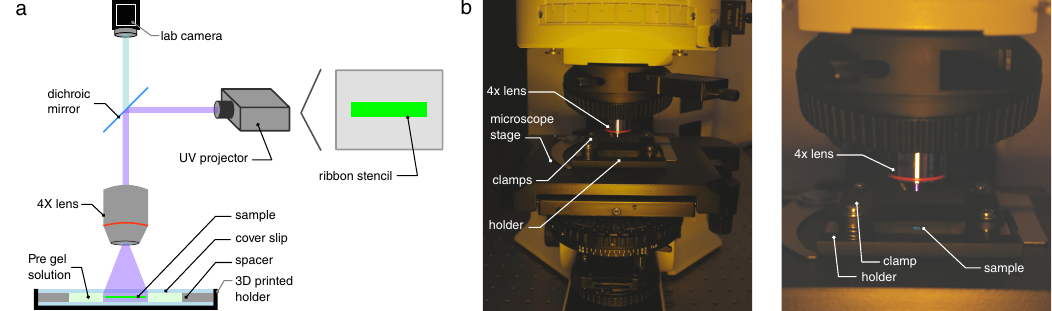}
    \caption[Experimental setup for the photolithographic synthesis of hydrogel samples.]{
    Experimental setup for the photolithographic synthesis of hydrogel samples.
    \begin{captionlist}
        \item \label{fig:polymerization_setup:sketch} A sketch of the optical path implemented in the upright microscope Olympus BX61.
        \item \label{fig:polymerization_setup:pictures} Pictures of the actual setup. 
    \end{captionlist}
    During fabrication, the pre-gel mixture, sandwiched between a glass slide and a cover slip distanced by a \qty{5}{\um} thick spacer, is exposed to UV light from a DLP projector. 
    The projected rectangular mask is focused through a 4X objective on the plane between the glass slide and cover slip. 
    The UV light activates the photoinitiator which, in turn, starts the polymerization and cross-linking reaction in the illuminated region.
    A dichroic mirror along the optical path allows to reflect the light from the projector to the sample and record images from the sample through a camera mounted on the microscope. The digital camera is used as a guide to focus and center the projected pattern. 
    The polymerization process of the hydrogel can be monitored from the digital camera via transmission illumination from below the sample with a UV-filtered light source.
    }
    \label{fig:polymerization_setup}
\end{figure*}

The pre-gel mixture, without photoinitiator, was degassed for \qty{20}{\minute} at a pressure of \qty{0.2}{bar}, while the stock solution of photoinitiator was prepared and added shortly before polymerization. 
To fabricate the samples, about \qty{50}{\ul} of pre-gel solution was placed on a \qty{1}{\mm} thick glass microscope slide and covered with a \qty{0.15}{\mm} thick cover slip separated from the glass slide by a \qty{5}{\um} thick spacer of AISI 301 stainless steel. 
The assembly was held together by a custom holder (fabricated in Onyx with a Markforged Mark Two 3D printer).
Polymerization was achieved by exposing for \qty{30}{\s} the mixture to a uniform pattern of UV light with peak wavelength of \qty{385}{\nm} and power density of \qty{650}{\uW\per\m\squared}. 
Specifically, a rectangular stencil was projected by means of a UV light DLP projector (Wintech PRO4500) mounted on an Olympus BX61 upright microscope (see \Fig\ref{fig:polymerization_setup}). 
The image from the projector was reflected by a suitable dichroic mirror and focused on the plane between the glass slide and cover slip through a Plan 4X objective (1-U2B222 from Olympus), forming a rectangle of dimensions $\qty{2460}{\um} \times \qty{214}{\um}$. A digital camera (acA4024-29uc from Basler) was used to focus the projected pattern and to image the hydrogel sample during fabrication.

After polymerization, the assembly was placed in  deionized (DI) water to remove unreacted residues and to detach the samples from the glass slide, then kept refrigerated at \qty{5}{\degreeCelsius}. 
All operations were performed at a temperature of \qty{21}{\degreeCelsius} in a room with UV-filtered light sources.
Before conducting the experiments, the samples were transferred in a \qty{1}{\g\per\L} solution of sodium chloride in DI water and allowed to attain equilibrium. 
In the experimental conditions, the gel samples had dimensions of $ \sim \qty{3.7} {\milli\m} \times \qty{0.39}{\milli\m}$ on the plane of the ribbon. 
Considering an average swelling factor of \num{1.66} from the nominal size of fabrication, we estimated $\sim \qty{8.3}{\um}$ for the sample thickness.

\begin{figure*}[t]
    \centering
    \includegraphics[width=\linewidth]{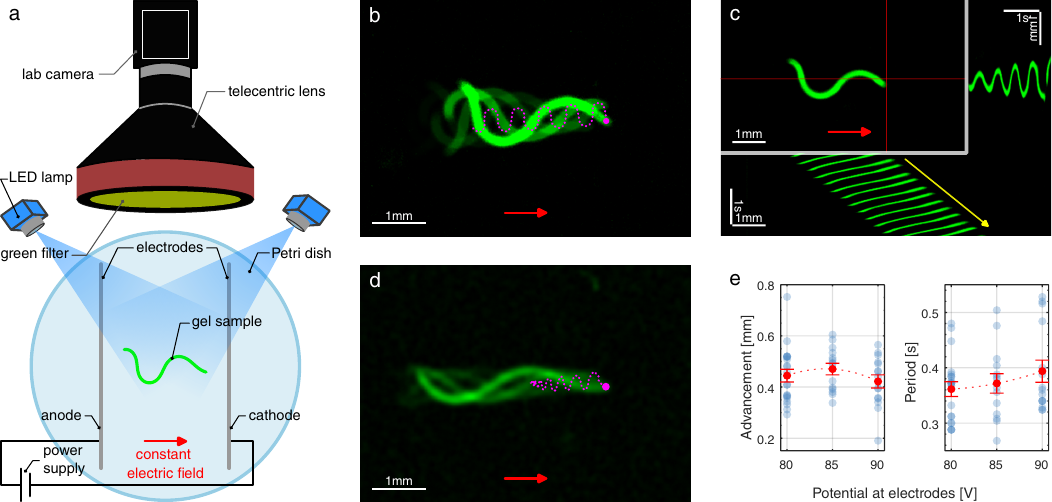}
    \caption[Experimental setup and observations.]{
    Experimental setup and observations.
    \begin{captionlist}
        \item \label{fig:exp:setup_sketch} Schematic illustration of the experimental setup used to explore the response of the hydrogel samples under a uniform and static electric field. 
        \item \label{fig:exp:filament_shape_change_ghosts} Gel filament swimming in the electrolytic solution towards the cathode by means of periodic, nonreciprocal shape changes. The direction and orientation of the electric field is represented by the red arrow.
        \item \label{fig:exp:kymographs} Kymographs of the swimmer shown in \ref{fig:exp:filament_shape_change_ghosts}) revealing the nonreciprocal character of the observed shape change. Specifically, the bottom quadrant represents the sections in time along the horizontal red line, while the right quadrant represents the sections along the vertical red line.
        \item \label{fig:exp:onset_ghotst} Onset of the periodic shape change, exhibiting the features of a Hopf bifurcation. 
        \item \label{fig:exp:experimental_data} Average advancement and period per cycle as measured from the experiments. Red dots represent the average of the measurements over all the trials at fixed electric potential, while error bars represent the standard error of the mean.
    \end{captionlist}
    }
    \label{fig:exp}
\end{figure*}

The experiments on the hydrogel samples were conducted in an electrolytic cell assembled inside a Petri dish of \qty{90}{\mm} in diameter.
Plane platinum electrodes measuring $\qty{50}{\mm}\times\qty{25}{\mm}$ were kept in place at a distance of $\qty{30}{\mm}$ by a custom designed 3D printed holder, see \Fig\ref{fig:exp}\ref{fig:exp:setup_sketch}. 
A constant potential difference, $\Delta V$, was applied between the electrodes from a programmable power supply (HMP4040 from Rohde \& Schwarz) inducing a uniform electric field orthogonal to the electrodes (neglecting boundary effects).
To perform the experiment, the cell was filled with about \qty{45}{\mL} of a \qty{1}{\g\per\L} solution of sodium chloride in DI water as electrolytic solution. 
The setup was lighted up by a pair of LED lamps with peak wavelength of \qty{455}{\nm} (ILH-OO01-DEBL-SC211 from ILS) and high contrast images were recorded at \qty{50}{fps} with a digital camera (ace acA4096-40uc from Basler) equipped with a telecentric lens and a high-pass filter (MVTC23024 and FGL515S from ThorLabs), taking advantage of the fluorescence of the samples.

The hydrogel samples were (almost) neutrally buoyant in the electrolytic solution, and were therefore free to rotate and translate. 
To set an initial configuration before application of the electric potential, the straight ribbons were manually oriented with the help of a worm pick to be horizontal and with their thinnest side facing the camera above the cell. 

\begin{figure*}[p]
    \centering
    \includegraphics[width=\linewidth]{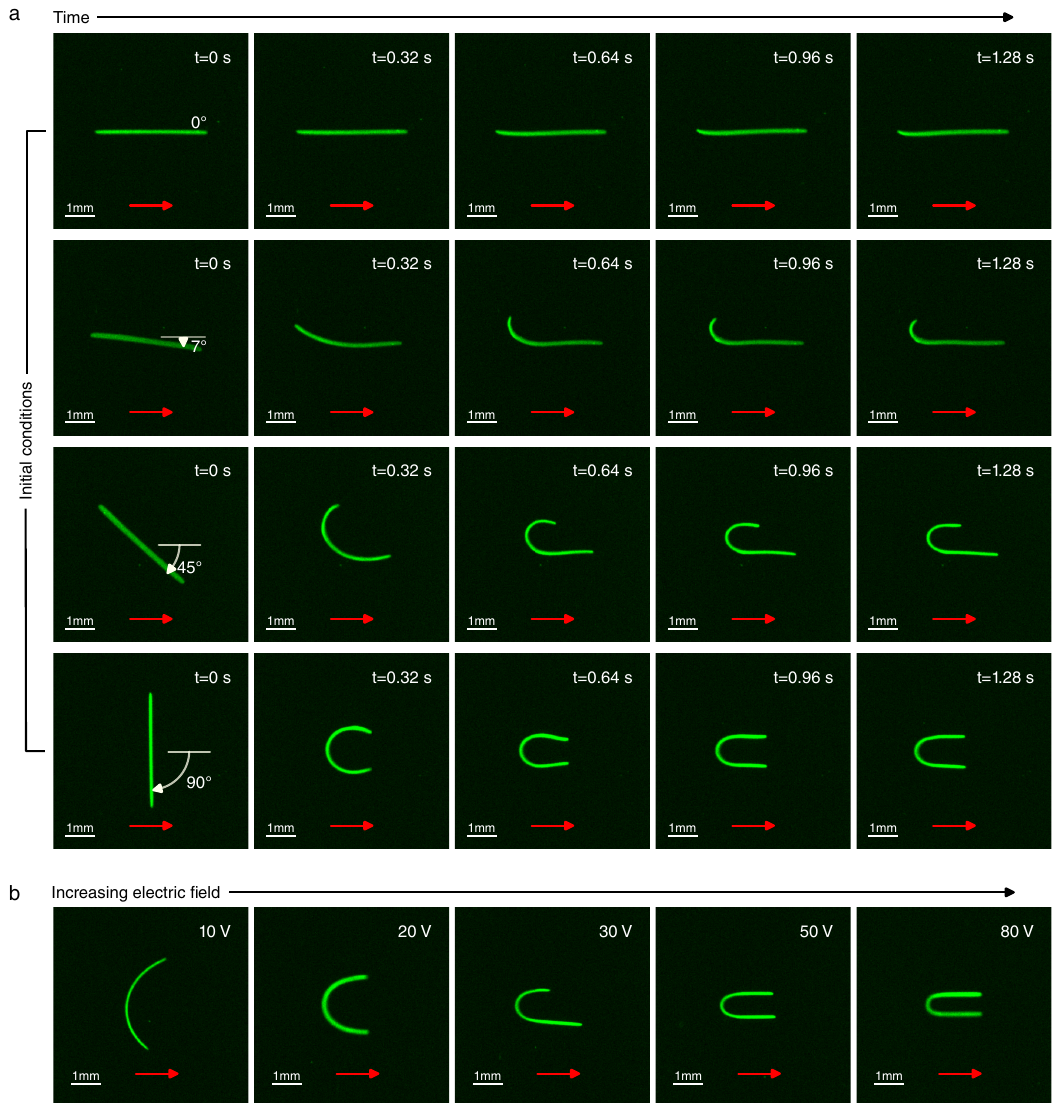}
    \caption[Active reconfiguration of PEH filaments in weak electric field.]{
    Active reconfiguration of PEH filaments in weak electric field.
    \begin{captionlist}
        \item \label{fig:exp_stat:dynamics} Dynamics from initial condition to equilibrium configuration. Time is measured from activation of the electrolytic cell with $\Delta V = \qty{40}{\volt}$. The direction of the electric field is highlighted by red arrows. In the first row, the filament barely changes shape, remaining aligned with the electric field. By introducing an angle in the initial condition, the filament evolves to an asymmetric configuration, as reported in the two central rows. In the last row, the filament evolves symmetrically to its equilibrium shape. The equilibrium shapes are characterized by a curvature distribution that is highest where the filament is orthogonal to the electric field and almost vanishing towards the ends. We remark that the maximum curvature is the same for all tests, except for the first row, in which the filament remains essentially straight. 
        \item \label{fig:exp_stat:voltage} Starting from a configuration (almost) orthogonal to the electric field, the filaments evolve to symmetric equilibrium configurations by actively bending towards the cathode, depending on the applied potential. These observations support the evolution law for the spontaneous curvature, as discussed in \sect\ref{MOD}.
    \end{captionlist}
    }
    \label{fig:exp_stat}
\end{figure*}

Upon activation by the electric field, different responses were observed. For values of $\Delta V \lesssim \qty{80}{V}$, the samples rapidly evolved to an equilibrium configuration by bending, about the softest axis of their cross-section, towards the cathode; the curvature field of such configurations depended on the relative orientation of the ribbon's axis to the electrodes and the intensity of the applied potential (\Fig \ref{fig:exp_stat} and \mov\ref{MOV:exp_stationary}). 

In sharp contrast, the hydrogel samples exhibited the emergence of periodic, oscillatory motion for higher values of the applied potential, as reported in \Fig\ref{fig:exp}\ref{fig:exp:filament_shape_change_ghosts} for the representative case of $\Delta V = \qty{85}{V}$.
Starting from an initial condition with the ribbon almost aligned to the electric field, such periodic oscillations allowed the hydrogel sample to swim towards the cathode (see \mov\ref{MOV:exp_swimming}), a fact that hints to the nonreciprocal nature of the observed shape change, in view of the constraints of low Reynolds number hydrodynamics. 
This is revealed by the kymographs reported in \Fig\ref{fig:exp}\ref{fig:exp:kymographs}: indeed, the shape change is characterized by a waveform that travels backward along the filament (inclined bands in the horizontal kymograph) with nonuniform velocity, which is highest in the central sector, and increasing in amplitude while traveling (vertical kymograph). 
One cycle of shape change comprises two traveling waves, which are symmetric about the horizontal axis and correspond to a pair of consecutive bands in the horizontal kymograph. 
The  extremities of such bands correspond to those of the filament during its motion, so that the inclination of the line connecting them allows to estimate the average translational velocity of the swimmer (yellow arrow).

Interestingly, the above described phenomenon emerges from a transient phase in which the filament departs from the initial, straight configuration through oscillations whose amplitude increases in time to eventually bring the system on a periodic orbit (\Fig\ref{fig:exp}\ref{fig:exp:onset_ghotst} and \mov \ref{MOV:exp_onset}). 
This behavior strongly resembles the features of a Hopf bifurcation~\cite{Strogatz2019}, which has been advocated to be responsible for the emergence of flagellar wave-like patterns~\cite{Camalet2000}.

To prove the robustness of the observed phenomenon and the repeatability of the procedure, experiments were carried out multiple times, on different hydrogel samples, and for different values of the electric potential, namely $\Delta V = \{80, 85, 90\}$\,V.
Specifically, the data reported in \Fig\ref{fig:exp}\ref{fig:exp:experimental_data} refers to 49 trials on 7 different samples.
For each experimental trial, we measured the net advancement per cycle and the period of the shape change, averaged over five cycles. 
These two quantities are significant features of the swimming dynamics, and can both be easily extracted from calibrated images (see~\ref{MANDM:image_processing}).
The results of the experimental campaign are summarized in the plots of \Fig\ref{fig:exp}\ref{fig:exp:experimental_data}, which also report  the average of the measurements (at different levels of applied potential) and the standard error of the mean. 
Broadly speaking, a weak dependence upon the electric potential was observed in the swimming behavior, for the considered range of $\Delta V$.
On the other hand, while swimming of the hydrogel samples was always observed, we remark here that some factors may have influenced our measurements, such as discrepancies between samples and disturbances from the environment.

Despite the symmetry of the experimental setup and the uniformity of the hydrogel samples, a striking feature of the observed phenomenon is that swimming always occurred along the electric field lines and in the direction of the cathode. 
This observation led us to speculate that the swimming direction of the samples could be controlled by reorienting the electric field, an hypothesis that will be investigated in the following.

\section{A minimal model for PEH ribbons}
\label{MOD}

\begin{figure*}
    \centering
    \includegraphics[width=\linewidth]{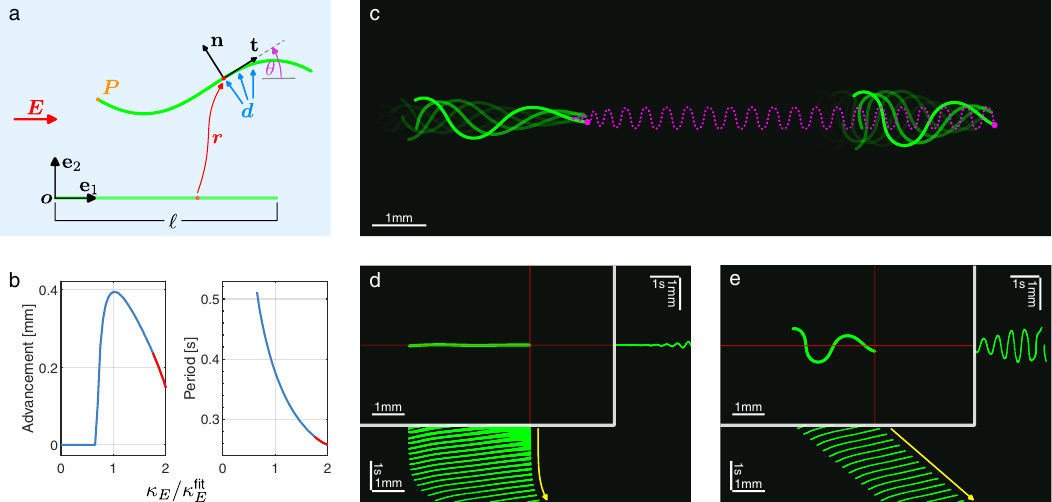}
    \caption[Mathematical model and numerical experiments.]{
    Mathematical model and numerical experiments. 
    \begin{captionlist}
        \item \label{fig:model:kinematics} 
        Kinematics of the planar swimming of a morphoelastic filament.
        \item \label{fig:model:kappaE_sweep} 
        Swimming descriptors (advancement per cycle and shape change period) for different values of $\kappa_E$.
        Notice that best swimming performance is obtained for $\kappa_E/\kappa_E\fit \simeq 1$. 
        The red parts in the curves correspond to shape changes characterized by self-intersection of the filament, and are hence not representative of the physical experiment.
        \item \label{fig:model:simulation_fit} 
        Swimming of an active filament as computed from the numerical solution of the nonlinear model equations. 
        The figure reports the evolution from the onset of the instability, characterized by growing oscillations, to the steady oscillatory solution.
        We report in magenta the trajectory of the filament's tip and in shades of green its configurations over one cycle of shape change near the onset of the dynamics and at the limit cycle. 
        \item \label{fig:model:kym_onset} Kymographs near the onset of swimming and 
        \item \label{fig:model:kym_limit} at the limit cycle.
    \end{captionlist}
    }
    \label{fig:model}
\end{figure*}

We introduce a mathematical model that will prove suitable to capture the physics behind the self-sustained oscillations and the swimming behavior of PEH ribbons, as observed in the experiments.
In the present model, the hydrogel ribbon is reduced to a planar, inextensible, morphoelastic filament \cite{Goriely2017} of length $\ell$, which is constrained to move in the plane identified by the orthonormal basis $(\eb{1},\eb{2})$, see \Fig\ref{fig:model}\ref{fig:model:kinematics}. 
In the following, $s \in (0, \ell)$ denotes the arc-length parameter of the filament's centerline and $t \in [0, +\infty)$ time.

To naturally treat the constraint of inextensibility, we introduce the angle, $\theta(s,t)$, between the line tangent to the filament and the horizontal axis $\eb{1}$.
Then, the unit tangent and the unit normal to the filament read $\dct{t}(s,t) = \cos\theta(s,t)\,\eb{1} + \sin\theta(s,t)\,\eb{2}$ and $\dct{n}(s,t) = - \sin\theta(s,t)\,\eb{1} + \cos\theta(s,t)\,\eb{2}$, respectively, such that the motion of the filament $(s,t) \mapsto \vct{r}(s,t)$ reads
\begin{equation}
    \label{MOD:eq:position}
    \vct{r}(s,t) = \vct{r}(0,t) + \int_{0}^{s} \dct{t}(\sigma,t) \diff\sigma. 
\end{equation}
Moreover, upon noting that $\partial_t \dct{t}(s,t) = \dct{n}(s,t)\partial_t \theta(s,t)$, the velocity of the filament follows as
\begin{equation}
	\label{MOD:eq:velocity}
	\vct{v}(s,t) = \partial_t\vct{r}(s,t) = \partial_t\vct{r}(0,t) + \int_{0}^{s} \dct{n}(\sigma,t) \partial_t \theta(\sigma,t) \diff \sigma.
\end{equation} 

Assuming the filament to be neutrally buoyant, the only external force acting on it comes from the hydrodynamic interactions with the surrounding fluid. 
To keep the model as simple as possible, such interactions are rendered through RFT \cite{Gray1955,Lighthill1976}, an approach that allows to represent the viscous drag force (per unit length) acting on a slender body at low Reynolds number via two resistive coefficients as:
\begin{equation}
    \label{MOD:eq:d_RRFT}
    \vct{d}(s,t) = \left[-\mu_{\parallel} (\dct{t}\otimes\dct{t}) - \mu_{\perp} (\dct{n}\otimes\dct{n})\right] \partial_t{\vct{r}}(s,t).
\end{equation}
Here, $\mu_{\parallel}$ and $\mu_{\perp}$ are the \emph{tangent} and the \emph{normal} resistive coefficients, respectively, which have been computed in~\cite{koens_2017} for a slender body of elliptic cross-section with semi-axes $a$ and $b$ in the limit of $a \ll b \ll \ell$
\begin{equation}
    \label{MOD:eq:resistive_coefficients}
    \mu_{\parallel} = \frac{4\pi\mu}{2\ln\left(2\ell/b\right) - 1},\quad
    \mu_{\perp} = \frac{4\pi\mu}{\ln\left(2\ell/b\right)},
\end{equation}
where $\mu$ is the dynamic viscosity of the surrounding fluid.
According to expressions \eqref{MOD:eq:velocity} and \eqref{MOD:eq:d_RRFT}, the drag force distribution over the filament is a function of $\theta(s,t)$, its time derivative $\partial_t \theta(s,t)$, and the velocity $\partial_t\vct{r}(0,t)$ of the extremity $s=0$.

We neglect inertia in setting the model, so that the momentum balance laws for the filament read~\cite{antman2005}
\begin{subequations}
    \label{MOD:eq:MBLs}
    \begin{align}
        \partial_s\vct{C}(s,t) + \vct{d}(s,t) &= \vct{0}, \label{MOD:eq:MBLs:BLM} \\[0.5mm]
        \partial_s\vct{M}(s,t) + \partial_s\vct{r}(s,t) \times \vct{C}(s,t) &= \vct{0}, \label{MOD:eq:MBLs:BAM}
    \end{align}
\end{subequations}
where $\vct{C}(s,t)$ and $\vct{M}(s,t)$ are the internal contact force and bending moment, respectively. 
The balance equations are complemented by suitable initial and boundary conditions; for the problem at hand, the latter read $\vct{C}(0,t) = \vct{C}(\ell,t) = \vct{0}$ and $\vct{M}(0,t) = \vct{M}(\ell,t) = \vct{0}$. 
In passing, we remark that such boundary conditions together with equations \eqref{MOD:eq:MBLs} correspond to the requirement that the total force and the total torque acting on the filament vanish. 

In the framework of morphoelasticity~\cite{Goriely2017}, the bending moment is constitutively prescribed through the difference between the \emph{visible} curvature of the filament, $\partial_s \theta(s,t)$, and its time-dependent \emph{spontaneous} curvature, $\kappa(s,t)$, so that
\begin{equation}
    \label{MOD:eq:bending_moment_constitutive}
    \vct{M}(s,t) = B \left[ \partial_s\theta(s,t) - \kappa(s,t) \right] \eb{3},
\end{equation}
where $B$ is the bending stiffness of the filament and $\eb{3} = \eb{1} \times \eb{2}$. 
Notice that, in the present context, the spontaneous curvature is the internal degree of freedom encoding the active response of the hydrogel ribbon to the external electric field, and is such that it corresponds to the visible curvature in the absence of external forces, when the filament is at equilibrium.

We describe the time-evolution of the spontaneous curvature by the following phenomenological law \cite{cicconofri_2023}
\begin{equation}
    \label{MOD:eq:spontaneous_curvature_evolution}
    \tau \partial_t \kappa(s,t) + \kappa(s,t) = - \kappa_{E} \sin\theta(s,t),
\end{equation}
which is reminiscent of the Kelvin-Voigt rheological model and captures, despite its simplicity, the main features of the active reconfigurations of the hydrogel samples. 
In response to the electric field $\vct{E} = E \,\eb{1}$, the filament attains in a characteristic time $\tau$ (see \Fig\ref{fig:exp_stat}\ref{fig:exp_stat:dynamics} and \mov\ref{MOV:exp_stationary}) a \emph{target} curvature $\kappa_E$ (see \Fig\ref{fig:exp_stat}\ref{fig:exp_stat:voltage}), whose value depends on that of $E$ and has concordant sign, as the hydrogel samples bend towards the cathode. 
In the physical experiments, the equilibrium curvature distribution depends on the orientation of the filament relative to the electric field, a feature that is captured by the model through the modulation by $\sin\theta(s,t)$ of the forcing term in the equation (see \Fig\ref{fig:exp_stat} and \mov\ref{MOV:exp_stationary}).

In preparation for the subsequent analysis, we now recast the model equations in scalar components. 
In particular, since the viscous drag force \eqref{MOD:eq:d_RRFT} is naturally expressed in tangential and normal components, we proceed by expressing the balance equations \eqref{MOD:eq:MBLs} in the local basis of $\{\dct{t},\dct{n}\}$. 
Hence, we decompose the contact force as $\vct{C}(s,t) = A(s,t) \dct{t}(s,t) + V(s,t) \dct{n}(s,t)$, where $A(s,t)$ is the axial component and $V(s,t)$ is the shearing component.
Having noted that $\partial_s\dct{t}(s,t) = \dct{n}(s,t)\partial_s\theta(s,t)$ and that $\partial_s\dct{n}(s,t) = -\dct{t}(s,t)\partial_s\theta(s,t)$, we obtain
\begin{equation}
    \label{MOD:eq:C_decomposition_derivative}
    \partial_s \vct{C}(s,t) = \left[\partial_s A(s,t) - V(s,t) \partial_s \theta(s,t)\right] \dct{t}(s,t) + \left[\partial_s V(s,t) + A(s,t) \partial_s \theta(s,t)\right]\dct{n}(s,t).
\end{equation}
Consequently, projection of \eqref{MOD:eq:MBLs:BLM} along the tangent and the normal unit vectors leads to
\begin{subequations}
	\label{MOD:eq:MBLs:BLM_components}
	\begin{align}
		\partial_s A(s,t) - V(s,t)\partial_s\theta(s,t) - \mu_{\parallel}v_{\parallel}(s,t) &= 0 ,\label{MOD:eq:MBLs:BLM_components:A} \\[0.5mm]
		\partial_s V(s,t) + A(s,t)\partial_s\theta(s,t) - \mu_{\perp}v_{\perp}(s,t) &= 0,  \label{MOD:eq:MBLs:BLM_components:V}
    \end{align}
\end{subequations}
where $v_{\parallel}(s,t) = \vct{v}(s,t)\cdot\dct{t}(s,t)$ and $v_{\perp}(s,t) = \vct{v}(s,t)\cdot\dct{n}(s,t)$ are the tangential and normal components of the velocity field, respectively. 
Regarding the balance of angular momentum \eqref{MOD:eq:MBLs:BAM}, this becomes
\begin{equation}
    \label{MOD:eq:MBLs:BAM_components}
    B\left[\partial_{s}^2\theta(s,t) - \partial_s\kappa(s,t)\right] + V(s,t) = 0 ,
\end{equation}
in light of the constitutive assumption \eqref{MOD:eq:bending_moment_constitutive} for the bending moment and of the relation $\partial_s r(s,t) = \dct{t}(s,t)$. 
Finally, the boundary conditions become
\begin{subequations}
	\label{MOD:eq:free_ends_BCs}
	\begin{alignat}{4}
		A(0,t) = V(0,t) 					&= 0,	&&	\qquad	&	A(\ell,t) = V(\ell,t) 						&= 0, \label{MOD:eq:free_ends_BCs:AV} \\[0.5mm]
		\partial_s\theta(0,t) - \kappa(0,t) &= 0,	&&	\qquad	&	\partial_s\theta(\ell,t) - \kappa(\ell,t) 	&= 0. \label{MOD:eq:free_ends_BCs:M}
	\end{alignat}
\end{subequations}

\section{Numerical experiments on morphoelastic swimming filaments}
\label{MOD:SIM}

To interpret the experimental observations on swimming PEH ribbons in light of the just presented mathematical model, we combine numerical simulations of the governing equations with their stability analysis, as detailed in the present and in the following sections. 

Numerical simulations were performed via the Finite Element discretization of the nonlinear model equations (see~\ref{NUM}). 
Interestingly, oscillations of the virtual filament could be achieved for $\kappa_E$ of sufficient magnitude, in analogy with the physical experiments. 
To quantitatively compare numerical results with physical experiments we proceeded as follows. 
Among all trials, we selected as representative experiment the one closest to the average values of the kinematic features shown in \Fig\ref{fig:exp}\ref{fig:exp:experimental_data}, which we found to correspond to $\Delta V = \qty{85}{V}$. 
The physical dimensions, $\{\ell, a, b\}$, of the relevant sample were measured from calibrated images (recall \sect\ref{EXP} and see~\ref{MANDM:image_processing}).
To retrieve the other model parameters, $\{B, \tau, \kappa_E\}$, we exploited a gradient-based optimization procedure (see~\ref{NUM:FIT}) to minimize a cost function defined as
\begin{equation}
	\label{NUM:FIT:eq:cost_fun}
	\mathcal{C} = \sum_{k=1}^3 \left(\frac{c_k^\textrm{e}}{c_k^\textrm{n}} - 1\right)^2 ,
\end{equation}
accounting for discrepancies between the experimental (superscript $\mathrm{e}$) and the numerical (superscript $\mathrm{n}$) results in terms of 
\begin{inlinenum}
	\item net advancement per cycle, $c_1$;
	\item period of shape change, $c_2$;
	\item lateral displacement of the filament's tip, $c_3$.
\end{inlinenum}

The model parameters determined by the optimization procedure are reported in \tab\ref{MOD:SIM:tab:parameters} and correspond to $\mathcal{C} \simeq \num{3.46E-2}$. 
The dynamic viscosity of the electrolytic solution in the experimental conditions was set to $\mu = \qty{8E-4}{\pascal\s}$, as determined via rheological measurements (see~\ref{MANDM:VISC}).

\begin{table}
    \centering
    \caption{Model parameters for the selected representative experiment as measured or determined by the optimization procedure.}
    \label{MOD:SIM:tab:parameters}
    \begin{tabular}{l S[table-format=1.2e-1] l}
        Measured and fitted model parameter             & \multicolumn{1}{c}{Value} & Units \\
        \midrule
        Filament's length, $\ell$                       & 3.70                      & \unit{\milli\meter} \\ 
        Filament's width, $2b$                           & 3.96E-1                   & \unit{\milli\meter} \\
        Filament's thickness, $2a$                       & 8.30E-3                   & \unit{\milli\meter} \\ 
        Filament's bending stiffness, $B\fit$           & 1.94E-8                   & \unit{\newton\milli\meter\squared} \\ 
        Characteristic relaxation time, $\tau\fit$      & 3.75E-1                   & \unit{\second} \\     
        Target spontaneous curvature, $\kappa_{E}\fit$  & 1.91E1                    & \unit{\per\milli\meter} \\
        \bottomrule
    \end{tabular}
\end{table}

We report in \Fig\ref{fig:model}\ref{fig:model:simulation_fit} the results from a numerical simulation carried out for the fitted model parameters of \tab\ref{MOD:SIM:tab:parameters} and assuming an initial condition with the filament straight and inclined by \qty{0.01}{\degree} with respect to $\eb{1}$, thus reproducing experimental conditions.
The motion of the morphoelastic filament is characterized by a  transient phase in which small amplitude oscillations emerge from the initial configuration and later evolve into a steady oscillatory motion (\mov\ref{MOV:swimmer_virtual}).
In striking similarity with the physical experiment, such periodic shape change allows the filament to self-propel in the viscous fluid along the horizontal axis, and is thus of nonreciprocal nature. 
This feature of the periodic, wave-like shape change is highlighted by the kymographs shown in \Fig\ref{fig:model}\ref{fig:model:kym_onset} and \Fig\ref{fig:model}\ref{fig:model:kym_limit}, corresponding, respectively, to the initial transient and the steady phase of the motion depicted in \Fig\ref{fig:model}\ref{fig:model:simulation_fit}. 
From the kymographs of \Fig\ref{fig:model}\ref{fig:model:kym_onset} we notice that the increase in the amplitude of the oscillations is accompanied by the increase in the average swimmer's velocity (yellow arrow). 
About the kymographs in \Fig\ref{fig:model}\ref{fig:model:kym_limit}, we underline the remarkable agreement with \Fig\ref{fig:exp}\ref{fig:exp:kymographs}, in support of the effectiveness of the proposed model in capturing all the fundamental features of the physical phenomenon at hand.

In the experiments, the occurrence of swimming critically depended on the intensity of the applied electric field. 
We explored this aspect by carrying out a parametric study in which all the model parameters, but the target curvature, were kept as in \tab\ref{MOD:SIM:tab:parameters}. 
In particular, we computed the effect of the target curvature upon the swimmer dynamics regarding the net advancement per cycle of shape change and its period. 
The results for $\kappa_E / \kappa_E\fit \in [0,2]$ are shown in \Fig\ref{fig:model}\ref{fig:model:kappaE_sweep} as obtained by running multiple simulations for distinct values of $\kappa_E$ and computing the relevant quantities by averaging the numerical results over 15 cycles of the periodic motion.
We notice that the swimmer's performance, in terms of net advancement per cycle, is best for $\kappa_E / \kappa_E\fit \simeq 1$. 
Both the net advancement and the period of shape change exhibit a dependence upon $\kappa_E$. 
Interestingly, a sharp transition in the response of the system occurs at a critical value of $\kappa_E\crit / \kappa_E\fit \simeq 0.65$, a finding that closely resembles the experimental observation that swimming of the PEH ribbons could only be realized for $\Delta V \gtrsim \qty{80}{V}$.

\begin{figure}[t]
    \centering
    \includegraphics[width=0.55\linewidth]{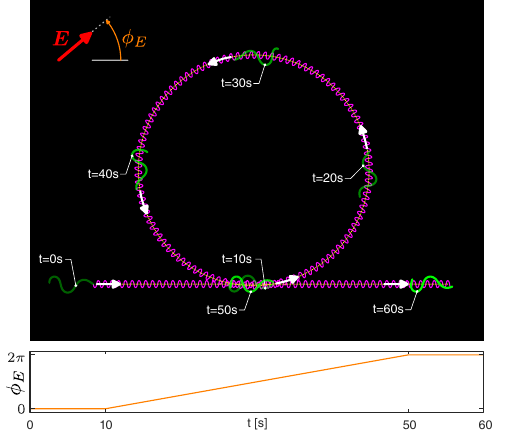}
    \caption{
    Control of the swimmer trajectory.
    By rotating the electric field as described by the time series in the lower panel, it is possible to drive the robotic swimmer along a circular trajectory. The simulation was performed for the model parameters reported in \tab\ref{MOD:SIM:tab:parameters}.
    }
    \label{fig:trajectories}
\end{figure}

We further exploit the proposed model to investigate by means of numerical simulations the possibility to control the trajectory of the swimmer by taking advantage of the \emph{electrotaxis} observed in the physical experiments. 
In our model, the current orientation of the filament in the electric field is taken into account in the forcing term at the right-hand side of \eqref{MOD:eq:spontaneous_curvature_evolution}; hence, changing the orientation of the electric field simply amounts to substituting the evolution law for the spontaneous curvature of \eqref{MOD:eq:spontaneous_curvature_evolution} with
\begin{equation}
    \tau\partial_t \kappa(s,t) + \kappa(s,t) = - \kappa_E \sin\left(\theta(s,t) - \phi_E(t)\right)
\end{equation}
where $\phi_E\left(t\right)$ is the inclination of the electric field relative to $\eb{1}$, positive when anticlockwise. 
We report in \Fig\ref{fig:trajectories} the results from a numerical simulation in which the orientation $\phi_E\left(t\right)$ of the electric field evolves in time with constant angular velocity, as shown in the lower panel of the figure. 
We notice that the swimmer's axis continuously aligns to the electric field, so that the resulting average trajectory (depicted in yellow in the upper panel of the figure) is circular, see also \mov\ref{MOV:circus}. 
Despite its simplicity, the presented example provides evidence of the possibility to steer the hydrogel swimmer in virtually any direction and hence to control its trajectory. 

\section{Swimming arises from flutter instability}
\label{LIN}

Intrigued by the transition to oscillatory motion observed in the numerical simulations (recall \Fig\ref{fig:model}\ref{fig:model:kappaE_sweep}), and its proximity to experimental observations, we investigated its origin via the stability analysis of the governing equations in relation to the relevant model parameters.

In this regard, a deeper insight on how the physical parameters control the behavior of the system can be obtained by looking at the governing equations in nondimensional form. 
We take $\ell$ as the characteristic length and $\tau$ as the characteristic time, such that $\bar{s} = s/\ell \in [0,1]$ and $\bar{t} = t / \tau \in [0,+\infty)$ are the dimensionless arc-length and time, respectively.
Furthermore, we normalize forces by $B\ell^{-2}$, velocities by $\ell\tau^{-1}$, and curvatures by $\ell^{-1}$. 

With a slight abuse of notation, we next use the symbols introduced for the physical quantities to denote their dimensionless counterpart and drop the overbar from dimensionless arc-length and time. 
Upon introducing the three dimensionless groups
\begin{equation}
    \label{MOD:eq:nondim_groups}
		\eta_{\parallel} = \frac{\mu_{\parallel}\ell^4}{B\tau}, 
		\qquad
		\eta_{\perp} = \frac{\mu_{\perp}\ell^4}{B\tau}, 
		\qquad
		\chi = {\kappa_{E}}{\ell}, 
\end{equation}
the balance equations \eqref{MOD:eq:MBLs:BLM_components} and \eqref{MOD:eq:MBLs:BAM_components} become
\begin{subequations}
	\label{MOD:eq:nondim_MBLs}
	\begin{align}
		\partial_s A(s,t) - V(s,t)\partial_s\theta(s,t) - \eta_{\parallel}v_{\parallel}(s,t) &= 0, \label{MOD:eq:nondim_MBLs:BLM:A} \\[0.5mm]
		\partial_s V(s,t) + A(s,t)\partial_s\theta(s,t) - \eta_{\perp}v_{\perp}(s,t) &= 0,  \label{MOD:eq:nondim_MBLs:BLM:V} \\[0.5mm]
        \partial_{s}^2\theta(s,t) - \partial_s\kappa(s,t) + V(s,t) &= 0,\label{MOD:eq:nondim_MBLs:BAM}
    \end{align}
\end{subequations}
whereas the evolution law \eqref{MOD:eq:spontaneous_curvature_evolution} yields 
\begin{equation}
    \label{MOD:eq:nondim_EVO}
    \partial_t\kappa(s,t) + \kappa + \chi\sin\theta(s,t) = 0.
\end{equation}
Finally, the boundary conditions \eqref{MOD:eq:free_ends_BCs} in nondimensional form become
\begin{subequations}
	\label{MOD:eq:nondim_free_ends_BCs}
	\begin{alignat}{4}
		A(0,t) = V(0,t) 					&= 0, &&\qquad	& 	A(1,t) = V(1,t) 						&= 0, \\[0.5mm]
		\partial_s\theta(0,t) - \kappa(0,t) &= 0, &&\qquad	& 	\partial_s\theta(1,t) - \kappa(1,t) 	&= 0.
	\end{alignat}
\end{subequations}
We remark that the parameter $\chi$ is a dimensionless measure of the target curvature. 
Instead, the parameters $\eta_\parallel$ and $\eta_\perp$ can be interpreted as the delay between the passive reconfiguration of the filament in the viscous fluid and its active response to the external stimulus. 
Such delay plays a key role for the existence of self-sustained, oscillatory solutions to the governing equations, as shown later.

The system of nondimensional equations \eqref{MOD:eq:nondim_MBLs}--\eqref{MOD:eq:nondim_free_ends_BCs} admits the trivial solution
\begin{equation}
	\label{LIN:eq:trivial_equilibrium}
		A^0(s,t) = 0, \quad V^0(s,t) = 0, \quad \theta^0(s,t) = 0, \quad \kappa^0(s,t) = 0, \quad v_\parallel^0(s,t) = 0, \quad v_\perp^0(s,t) = 0,
\end{equation}
corresponding to a straight filament at rest and aligned with the external stimulus.
To linearize the governing equations around this trivial solution, we assume asymptotic expansions of the relevant fields in the small parameter $\epsilon$. 
To begin with, we write
\begin{equation}
	\label{LIN:eq:linearized_fields}
		\theta(s,t) = \theta^0 + \epsilon\, \theta^1(s,t), \quad
        v_\parallel(s,t) = v_\parallel^0 + \epsilon\, v_\parallel^1(s,t), \quad
        v_\perp(s,t) = v_\perp^0 + \epsilon\, v_\perp^1(s,t),
\end{equation}
so that retaining first order terms in $\epsilon$ after substitution in \eqref{MOD:eq:velocity} leads to 
\begin{equation}
	\label{LIN:eq:velocity_local_components}
		v_{\parallel}^1(s,t) = \partial_t r_1^1(0,t),\
		\qquad
		v_{\perp}^1(s,t) = \partial_t r_2^1(s,t),
\end{equation}
where $r_1^1(s,t)$ and $r_2^1(s,t)$ are the first order perturbations of the Cartesian coordinates of $\vct{r}(s,t)$ in the reference frame $\{\eb{1},\eb{2}\}$, recall the sketch of \Fig\ref{fig:model}\ref{fig:model:kinematics}.  

Next, we write
\begin{equation}
	\label{LIN:eq:linearized_fields_B}
		A(s,t) = A^0 + \epsilon\, A^1(s,t), \quad
		V(s,t) = V^0 + \epsilon\, V^1(s,t), \quad
		\kappa(s,t) = \kappa^0 + \epsilon\, \kappa^1(s,t),
\end{equation}
and plug such perturbations into \eqref{MOD:eq:nondim_MBLs}--\eqref{MOD:eq:nondim_free_ends_BCs}.
Noting from \eqref{MOD:eq:position} that $\theta^1(s,t) = \partial_s r_2^1(s,t)$, we finally obtain the linearized governing equations
\begin{subequations}
	\label{LIN:eq:MBLs}
	\begin{align}
		\partial_s A^1(s,t) - \eta_{\parallel}\partial_t r_1^1(0,t) &= 0, \label{LIN:eq:MBLs:BLM:A} \\[0.5mm]
		\partial_s V^1(s,t) - \eta_{\perp}\partial_t r_2^1(s,t) &= 0, \label{LIN:eq:MBLs:BLM:V} \\[0.5mm]
		\partial_s^3r_2^1(s,t)  - \partial_s\kappa^1(s,t) + V^1(s,t) &= 0, \label{LIN:eq:MBLs:BAM} \\[0.5mm]
		\partial_t\kappa^1(s,t) + \kappa^1(s,t) + \chi\, \partial_s r_2^1(s,t) &= 0 .\label{LIN:eq:MBLs:EVO}
	\end{align}
\end{subequations}
Accounting once again for the relation $\theta^1(s,t) = \partial_s r_2^1(s,t)$, the linearized boundary conditions directly follow from \eqref{MOD:eq:nondim_free_ends_BCs} and are here reported for completeness
\begin{subequations}
	\label{LIN:eq:lin_BCs_A}
	\begin{alignat}{4}
		A^1(0,t) = V^1(0,t) 					&= 0,	&& \qquad	& A^1(1,t) = V^1(1,t) 						&= 0, \label{LIN:eq:lin_BCs_A:forces}\\[0.5mm]
		\partial^2_s r_2^1(0,t) - \kappa^1(0,t) &= 0,	&& \qquad	& \partial^2_s r_2^1(1,t) - \kappa^1(1,t) 	&= 0.\label{LIN:eq:lin_BCs_A:moment}
	\end{alignat}
\end{subequations}
Now, by integrating \eqref{LIN:eq:MBLs:BLM:A} and imposing the boundary conditions \eqref{LIN:eq:lin_BCs_A:forces} relevant to the incremental axial force, we get $A^1(s,t) = 0$ and equivalently $\partial_t r_1^1(0,t) = 0$. 
Hence, at leading order, the filament motion is null in the direction of the external stimulus.
This reflects the experimental and computational observation about the progressive increase of the swimmer's average velocity at the onset of flutter instability. 
Also, notice that in the linear setting, having removed \eqref{LIN:eq:MBLs:BLM:A}, the system dynamics is controlled by only two dimensionless groups, namely $\chi$ and $\eta_\perp$.

Combining equations \eqref{LIN:eq:MBLs:BLM:V} and \eqref{LIN:eq:MBLs:BAM}, we arrive at a system of two coupled and linear PDEs for the incremental lateral displacement and spontaneous curvature
\begin{subequations}
	\label{LIN:eq:lin_system}
	\begin{align}
		\partial_s^4 r_2^1(s,t)  - \partial_s^2\kappa^1(s,t) + \eta_{\perp}\partial_t r_2^1(s,t) &= 0, \label{LIN:eq:lin_system:theta} \\[0.5mm]
		\partial_t\kappa^1(s,t) + \kappa^1(s,t) + \chi\,\partial_s r_2^1(s,t) &= 0 . \label{LIN:eq:lin_system:kappa}
	\end{align}
\end{subequations}
As for the boundary conditions, these reduce to 
\begin{equation}
	\label{LIN:eq:lin_BCs_B}
	\begin{alignedat}{4}
		\partial_s^3 r_2^1(0,t) - \partial_s\kappa^1(0,t) &= 0,	&& \qquad	&	\partial_s^3 r_2^1(1,t) - \partial_s\kappa^1(1,t) &= 0, \\[0.5mm]
		\partial_s^2 r_2^1(0,t) - \kappa^1(0,t) &= 0,	&& \qquad	&	\partial_s^2 r_2^1(1,t) - \kappa^1(1,t) &= 0,
	\end{alignedat}
\end{equation}
and stand for zero shear force and zero moment at the filament's extremities.

Seeking solutions in time-harmonic form
\begin{equation}
	\label{LIN:eq:time_harmonic_ansatz}
		r_2^1(s,t) = \hat{r}_2(s) \ee^{\omega t}, \qquad
		\kappa^1(s,t) = \hat{\kappa}(s)\ee^{\omega t},
\end{equation} 
the linearized system of governing equations, together with relevant boundary conditions, leads to the eigenvalue problem
\begin{equation}
	\label{LIN:eq:eigenvalue_problem}
    \begin{bmatrix}
        -\frac{1}{\eta_\perp}\partial_s^4    &  \frac{1}{\eta_\perp} \partial_s^2 \\[1mm]
        -\frac{1}{\chi}\partial_s           &   -\frac{1}{\chi}
    \end{bmatrix}
    \begin{bmatrix}
        \hat{r}_2(s) \\[1mm]
        \hat{\kappa}(s)
    \end{bmatrix}
    = \omega 
    \begin{bmatrix}
        \hat{r}_2(s) \\[1mm]
        \hat{\kappa}(s)
    \end{bmatrix}.
\end{equation}
We proceed by plugging \eqref{LIN:eq:time_harmonic_ansatz} into \eqref{LIN:eq:lin_system:kappa}, which gives the relation
\begin{equation}
	\label{LIN:eq:kappa_r2_harmonic}
    \hat{\kappa}(s) = - \frac{\chi}{1+\omega}\,\partial_s\hat{r}_2(s),
\end{equation}
such that substitution of \eqref{LIN:eq:kappa_r2_harmonic} into \eqref{LIN:eq:lin_system:theta} leads to the following fourth-order ODE for the spatial-dependent incremental lateral displacement
\begin{equation}
	\label{LIN:eq:lin_ODE}
	\partial_s^4\hat{r}_2(s) + \frac{\chi}{1 + \omega}\, \partial_s^3\hat{r}_2(s) + \omega\,\eta_{\perp}\hat{r}_2(s) = 0 ,
\end{equation}
whose general integral is
\begin{equation}
	\label{LIN:eq:lin_ODE_sol}
	\hat{r}_2(s) = c_1 \ee^{\lambda_1 s} + c_2 \ee^{\lambda_2 s} + c_3 \ee^{\lambda_3 s} +  c_4 \ee^{\lambda_4 s},
\end{equation}
where $\lambda_i$, with $i = \{1,2,3,4\}$, are the complex roots of the quartic polynomial
\begin{equation}
	\label{LIN:eq:lin_ODE_characteristic_equation}
	\lambda^4 + \frac{\chi}{\omega+1}\, \lambda^3 + \eta_{\perp} \omega.
\end{equation}

Imposing the boundary conditions \eqref{LIN:eq:lin_BCs_B} to the general solution \eqref{LIN:eq:lin_ODE_sol} provides the characteristic equation for the circular frequency $\omega$ as a function of the dimensionless groups $\chi$ and $\eta_\perp$ 
\begin{equation}
	\label{LIN:eq:char_eq}
	f(\omega\,;\chi,\eta_\perp) = 0,
\end{equation}
which is explicitly reported in \ref{LIN:DERIVATION}.
Due to the transcendental nature of the characteristic equation \eqref{LIN:eq:char_eq}, closed form solutions are not available, and we resort to solving it numerically, as detailed in~\ref{LIN:ALG}. 

Denoting by $\omega_{\max} = \alpha + \ii \beta$ the eigenvalue with greatest real part, we found two possible scenarios depending on the dimensionless parameters: 
\begin{inlinenum}
    \item $\alpha<0$, such that the trivial equilibrium is stable and any perturbation fades away, and
    \item $\alpha > 0$ with $\beta \neq 0$, such that any perturbation of the equilibrium develops into oscillations of increasing amplitude. 
\end{inlinenum}
Here we refer to the phenomenon associated with the latter case as \emph{flutter instability}~\cite{Ziegler1977,bigoni2023flutter}.
As revealed by the nonlinear simulations, after this transient, the system evolves towards a limit cycle; thus, the transition from stability to flutter described here is an instance of \emph{supercritical Hopf bifurcation}~\cite{Strogatz2019}. 

\begin{figure*}[t]
    \centering
    \includegraphics[width=\linewidth]{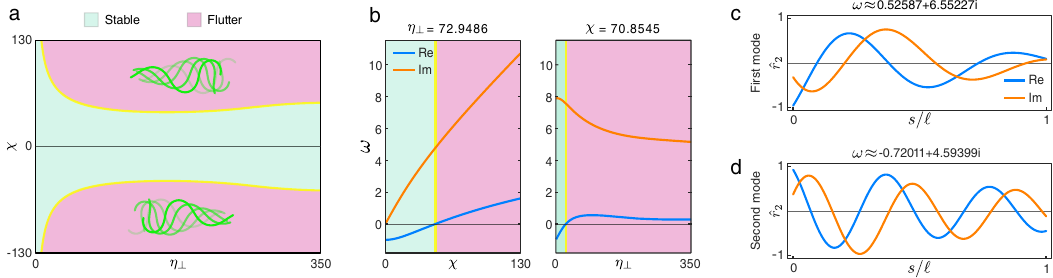}
    \caption[Linear stability analysis.]{
    Results of the linear stability analysis. 
    \begin{captionlist}
        \item \label{fig:stability:stability_map} Stability map in the $\eta_{\perp} - \chi$ parameters plane. 
        In the flutter regions, the swimmer is sketched according to the swimming direction as determined by the sign of $\chi$.
        \item \label{fig:stability:bifurcation_plot} Bifurcation plots varying one parameter at a time, either $\chi$ or $\eta_\perp$.
        \item\label{fig:stability:mode_1} First and \item\label{fig:stability:mode_2} second eigenmode of the fluttering swimmer, corresponding to the eigenvalue with largest and second largest real part, respectively, for the choice of model parameters. The second eigenvalue has negative real part and the two modes are separated by a mode with null corresponding eigenvalue, that is an equilibrium solution.
    \end{captionlist}
    }
    \label{fig:stability}
\end{figure*}

The stability map of the system is shown in \Fig\ref{fig:stability}\ref{fig:stability:stability_map} for $\eta_\perp \in [0,350]$ and $\chi \in [-130,130]$. 
At each value of $\eta_\perp$, we identify the critical value of the forcing parameter, $\chi$, at which the system transitions from stable to unstable.
We notice the symmetry of the map with respect to the horizontal axis, a feature reflecting the fact that the stability of the system is independent of the sign of $\chi$, which instead controls the swimming direction.
For the model parameters summarized in \tab\ref{MOD:SIM:tab:parameters}, we show in \Fig\ref{fig:stability}\ref{fig:stability:bifurcation_plot} the evolution of the real (blue curve) and of the imaginary (orange curve) part of the leading eigenvalue as a function of $\chi$ for fixed $\eta_\perp$ and vice-versa. 
Notice that oscillatory solutions always exist for $\chi \neq 0$, but they are damped by the viscous drag for $\chi$ below a critical value.
We also notice the relevance of $\eta_\perp$ in the occurrence of flutter. Indeed, the critical threshold increases and exits the range considered in our analysis, for small values of such parameter.
Finally, the first and the second eigenmodes of the filament are reported in \Fig\ref{fig:stability}\ref{fig:stability:mode_1} and in \Fig\ref{fig:stability}\ref{fig:stability:mode_2}, respectively. 
We remark the similarity between the first eigenmode and the configurations observed in the experiments of swimming PEH ribbons. 
We also underline that the real (blue curve) and the imaginary (orange curve) part of the eigenmodes are not proportional, such that their modulation in time  explains the nonreciprocal nature of the periodic shape changes observed in the experiments and in the numerical simulations.

\section{Discussion}
\label{CONC}

Our study provides experimental evidence that swimming of hydrogel-based soft robots can be achieved under minimal untethered actuation, that is, under a uniform and static electric field. 
This behavior has not been reported before and is here explained, via a minimal mathematical model, as an instance of flutter instability arising from the interplay between the active and the passive reconfigurations of the hydrogel robot in the fluid environment. 

By harnessing flutter instability, lower-level control tasks are embodied into the system, thus reducing the need for complex external actuation.
The swimmer, which only consists in a hydrogel ribbon, is capable to extract energy from a constant and uniform environmental stimulus, fully taking advantage of the internal activity of the constitutive material with remarkable performance. 
Interestingly, directed swimming of the soft robot is shown to occur along the electric field vector, a feature that permits to control its trajectory by simply reorienting the external stimulus, as proven by means of numerical simulations.

Despite its simplicity, the proposed mathematical model is descriptive of the most significant features of the swimming behavior observed in the experiments, thus allowing to shed light into the underlying physics. 
Considering its descriptive power, the model could be exploited to predict the behavior of the system across a range of material and geometric parameters and can hence be intended as a design tool for soft robots based on the same working principle. 
In this regard, the inherent geometric simplicity of the proposed model swimmer makes it suitable for miniaturization by means of advanced micro-fabrication techniques, such as two-photon polymerization lithography.
More generally, our findings contribute to tackle current challenges in soft robotics for the untethered control of biomimetic machines and timely respond to the need for design principles to achieve complex functionalities in active synthetic structures through minimal control. 
We believe that the design principle we have introduced for biomimetic structures based on PEH may find applications in a wider range of soft active materials, such as stimuli-responsive liquid crystal elastomers.

\section*{Aknowledgements}
\noindent
This work was supported by the Italian Ministry of University and Research (MUR) through the grants PRIN 2022 n.\,2022NNTZNM `DISCOVER' and `Dipartimenti di Eccellenza 2023-2027 (Mathematics Area)'. 
The experimental activities were carried out at the SAMBA laboratory of SISSA. 
G.N. and A.S.B. are members of the `Gruppo Nazionale di Fisica Matematica' (GNFM) of the `Istituto Nazionale di Alta Matematica' (INdAM).

\appendix

\section{Experimental materials and methods}
\label{MANDM}

\subsection{Image processing of swimming  hydrogel samples}
\label{MANDM:image_processing}

Experiments on polyelectrolyte hydrogel samples were conducted as described in \sect\ref{EXP}. 
In particular, image stacks were recorded at a frame rate of \qty{50}{fps} for hydrogel samples swimming straight towards the cathode and across the center of the electrolytic cell. 
The data set used in this work was collected from experiments executed during three days (August 8-10, 2023).
Experiments corresponding to \Fig\ref{fig:exp}\ref{fig:exp:onset_ghotst}, \Fig\ref{fig:exp_stat}, \mov\ref{MOV:exp_stationary} and \mov\ref{MOV:exp_onset} were executed during different sessions (August 23-24, 2023).

The recorded image stacks were calibrated by means of standard procedures and were then analyzed as follows:
\begin{enumerate}
    \item \label{IMAG:item:dataset} 
    For every successful experimental trial, we extracted from the relevant image stack the time series of images corresponding to 5 cycles of periodic shape change of the swimming filament. 
    We collected the time series from a total of 49 experimental trials carried out on 7 distinct hydrogel samples. 
    This data set was analyzed to produce \Fig\ref{fig:exp}\ref{fig:exp:experimental_data}. 
    In particular, the average period of the shape change over 5 cycles was determined from the number of images comprising the time series and tacking into account the frame rate of the digital camera. 
    As regards the average net advancement of the swimming samples, this was measured by manually tracking the initial and the final position of the sample's tip in the time series and dividing the distance between those two positions by the number of cycles.
    \item \label{IMAG:item:imagej} 
    Among all the trials, we selected as representative experiment the one closest to the average values of the kinematic descriptors reported in \Fig\ref{fig:exp}\ref{fig:exp:experimental_data}. 
    For such representative experiment, the motion of the swimming filament was entirely tracked by using the plugin \textsf{JFilament} \cite{smith_2010} of the open source image processing package \textsf{Fiji}. 
    The kinematic information extracted by tracking the motion of the filament was then used to retrieve the model parameters as described in \ref{NUM:FIT}. 
    The corresponding trajectory of the filament's tip is reported in \Fig\ref{fig:exp}\ref{fig:exp:filament_shape_change_ghosts}. 
    The same procedure was exploited for the analysis of the time series from an additional experimental trial to track the motion of a filament exhibiting the onset of the periodic shape change starting from an equilibrium configuration. 
    As for the previous case, the trajectory of the filament's tip is reported in \Fig\ref{fig:exp}\ref{fig:exp:onset_ghotst}.
\end{enumerate}

\subsection{Measurement of electrolytic solution viscosity}
\label{MANDM:VISC}

As detailed in \sect\ref{EXP}, the experiments on swimming hydrogel ribbons were carried out under LED light illumination to exploit the fluorescence of the  samples to record high contrast images for subsequent analysis. 
Illumination of the setup resulted in an increase of the temperature of the electrolytic solution from the ambient temperature to about \qty{32}{\degreeCelsius}, as measured in representative tests.
As the dynamic viscosity of the electrolytic solution, $\mu$, is a relevant physical parameter in the mathematical model, this was characterized by means of rheological tests performed with a MCR\,702e MultiDrive rheometer from Anton Paar equipped with a double gap measuring system DG26.7/T200/SS, suitable for low viscosity fluids.
Specifically, the measurements were carried out in rotational mode with shearing rates ranging from \qty{200}{\per\second} to \qty{600}{\per\second} and under controlled temperature from \qty{21}{\degreeCelsius} to \qty{40}{\degreeCelsius}.
An average dynamic viscosity of $\mu\approx\qty{8.031e-1}{\milli\pascal\second}$ was obtained for the representative temperature of \qty{32}{\degreeCelsius}.

\section{Numerical implementation}
\label{NUM}

The numerical experiments presented in \sect\ref{MOD:SIM} were performed by means of Finite Element simulations implemented in the commercial software \textsf{COMSOL Multiphysics~6.1 (build 357)}.

The formulation of the Finite Element model requires to recast the governing equations in weak form.
We introduce the symbols $\hat{A}$ and $\hat{V}$ to denote the test functions for the axial force, $A$, and for the shearing force, $V$, respectively. 
Then, multiplication of \eqref{MOD:eq:MBLs:BLM_components:A} by $\hat{A}$, of \eqref{MOD:eq:MBLs:BLM_components:V} by $\hat{V}$, and integration in space over the interval $[0,\ell]$ leads to the corresponding equilibrium of forces in weak form 
\begin{equation}
    \label{MOD:WEAK:eq:BLM_weak}
    \begin{aligned}
        \int_{0}^{\ell} [\partial_s A(\sigma,t) - V(\sigma,t)\partial_s\theta(\sigma,t) - \mu_{\parallel}v_{\parallel}(\sigma,t) ] \hat{A}(\sigma) \diff \sigma &= 0,\qquad \forall \, \hat{A}(s),\\[1mm]
        \int_{0}^{\ell} [\partial_s V(\sigma,t) + A(\sigma,t)\partial_s\theta(\sigma,t) - \mu_{\perp}v_{\perp}(\sigma,t) ] \hat{V}(\sigma) \diff \sigma &= 0,\qquad \forall \, \hat{V}(s),
    \end{aligned}
\end{equation}
where we account for the essential boundary conditions of \eqref{MOD:eq:free_ends_BCs:AV} at $s=0$, hence $\hat{A}(0) = \hat{V}(0) = 0$.
To also fulfill the boundary conditions \eqref{MOD:eq:free_ends_BCs:AV} at $s=\ell$,  we require that the total drag force acting on the filament vanishes, that is 
\begin{equation}
    \label{MOD:WEAK:eq:global_eq}
    \int_{0}^{\ell} [-\mu_{\parallel}v_{\parallel}(\sigma,t)\dct{t}(\sigma,t) -\mu_{\perp}v_{\perp}(\sigma,t)\dct{n}(\sigma,t) ] \diff \sigma = \vct{0}.
\end{equation}
This additional equation, coupled with suitable initial conditions, allows to determine the translational degree of freedom $\vct{r}(0,t)$ appearing in \eqref{MOD:eq:position} and to compute the configuration of the filament via integration of the tangent vector field.

Likewise, we denote by $\hat{\theta}$ the test function for the angular field $\theta$. The weak form of the equilibrium of moments is then obtained by testing \eqref{MOD:eq:MBLs:BAM_components} by $\hat{\theta}$, such that integration over $[0,\ell]$ and integration by parts, accounting for the natural boundary conditions \eqref{MOD:eq:free_ends_BCs:M}, leads to
\begin{equation}
	\int_{0}^{\ell} B\left[\partial_s\theta(\sigma,t) - \kappa(\sigma,t)\right]\partial_s\hat{\theta}(\sigma) + V(\sigma,t)\hat{\theta}(\sigma) \diff \sigma = 0,\qquad \forall\,\hat{\theta}(s). 
\end{equation}

Finally, the weak form of the evolution law \eqref{MOD:eq:spontaneous_curvature_evolution} is obtained by multiplying it by the corresponding test function $\hat{\kappa}$ and integrating over $[0,\ell]$, that is
\begin{equation}
	\int_{0}^{\ell} \left[\tau\partial_t\kappa(\sigma,t) + \kappa(\sigma,t) + \kappa_E\sin\theta(\sigma,t)  \right]\hat{\kappa}(\sigma) \diff \sigma = 0,\qquad\forall\,\hat{\kappa}(s).
\end{equation}

As for the discretization, the morphoelastic filament was divided into 30 uniformly spaced elements. 
On this mesh, the unknown fields were approximated by Lagrange shape function of degree 2 for the contact forces $A$ and $V$, of degree 3 for the spontaneous curvature $\kappa$, and of degree 4 for the kinematic descriptor $\theta$. 
The translational degree of freedom $\vct{r}(0,t)$ was determined by imposing the vanishing of total drag force, as discussed above. 
Thus, the configuration $\vct{r}(s,t)$ of the filament was computed from the tangent vector field using the built-in integration operator of order 4. 
Finally, integration in time was performed via the adaptive backward differentiation formula as time stepping method by setting a maximum time-step of \textsf{$\mathsf{\Delta t=\,} \num{1e-3}$}\,\unit{\second}.

\subsection{Optimization of the model parameters}
\label{NUM:FIT}
The model parameters that could not be directly measured, namely $\{B,\tau,\kappa_E\}$, were determined by means of an optimization procedure, minimizing the difference between results from numerical simulations and experimental measurements, as discussed in \sect\ref{MOD:SIM}.
The data for fitting was extracted from experimental images as described in~\ref{MANDM:image_processing}.
Simulations of the mathematical model were performed as described above. 
The optimization procedure was implemented in \textsc{Matlab} \textsf{R2022b} by using the \textsf{Optimization Toolbox}. To interface the optimization algorithm with \textsf{COMSOL} we took advantage of the \textsf{LiveLink for} \textsc{Matlab} module. In particular, minimization of the cost function \eqref{NUM:FIT:eq:cost_fun} was performed using \textsc{Matlab}'s \textsf{fmincon} solver. 

Evaluation of the cost function \eqref{NUM:FIT:eq:cost_fun} amounts to running a numerical simulation of the nonlinear model over a time interval adequate to reach a quasi-periodic solution to then extract the three kinematic quantities $\{c_1^{\mathrm{n}},c_2^{\mathrm{n}},c_3^{\mathrm{n}}\}$ for comparison with the experimental measurements $\{c_1^{\mathrm{e}},c_2^{\mathrm{e}},c_3^{\mathrm{e}}\}$.

An initial guess of the parameters $\{B^0, \tau^0, \kappa_E^0\}$ as in \tab\ref{SI:tab:optimization_parameters} was taken, based on qualitative analysis of the numerical simulations.
Optimization was then performed on three dimensionless and normalized parameters, namely $\{\alpha,\beta,\gamma\} \in [0.5,1.5]$, such that at each iteration $\kappa_E = \alpha\kappa_E^0$, $B = \beta B^0$, and $\tau = \gamma\tau^0$.

At each step of the optimization procedure, the parameters are updated by \textsf{fmincon} using a SQP algorithm: the gradient of the cost function is estimated and parameters updates are performed accordingly.
The algorithm stops when no significant gain can be obtained in an iteration. 
The final values of the model parameters after optimization are reported in \tab\ref{SI:tab:optimization_parameters} and correspond to $\mathcal{C}\simeq \num{3.46E-2}$.

\begin{table}
	\centering
	\caption{Model parameters for optimization procedure: initial guess and fitted parameters are reported.}
	\label{SI:tab:optimization_parameters}
	\begin{tabular}{l S[table-format=1.2e-1] S[table-format=1.2e-1] l}
        Parameter 							&
		\multicolumn{1}{c}{Initial guess}	& 
		\multicolumn{1}{c}{Fitted value}	& 
		Units \\
        \midrule
        Filament's bending stiffness, $B$           &	2.00E-8		& 1.94E-8	& \unit{\newton\milli\meter\squared} \\ 
        Characteristic relaxation time, $\tau$      &	3.00E-1		& 3.75E-1	& \unit{\second} \\     
        Target spontaneous curvature, $\kappa_{E}$  &	1.80E1		& 1.91E1   	& \unit{\per\milli\meter} \\
        \bottomrule
	\end{tabular}
\end{table}

\section{Details on the linear stability analysis}
We report explicitly the characteristic equation \eqref{LIN:eq:char_eq} and its derivation, along with the algorithm for the construction of the stability map depicted in \Fig\ref{fig:stability}\ref{fig:stability:stability_map}.

\subsection{Derivation of the characteristic equation}
\label{LIN:DERIVATION}
Imposing the boundary conditions \eqref{LIN:eq:lin_BCs_B} to the general solution \eqref{LIN:eq:lin_ODE_sol} yields the following linear and homogeneous system in the four constants of integration $\{c_1, c_2, c_3, c_4\}$
\begin{equation}
	\label{LIN:eq:constr_matrix}
	\begin{bmatrix}
		\displaystyle\lambda_1 \frac{\chi}{\omega + 1} + \lambda_1^2 & 
		\displaystyle\lambda_2 \frac{\chi}{\omega + 1} + \lambda_2^2 & 
		\displaystyle\lambda_3 \frac{\chi}{\omega + 1} + \lambda_3^2 & 
		\displaystyle\lambda_4 \frac{\chi}{\omega + 1} + \lambda_4^2 \\[3mm]
		\displaystyle\lambda_1^2 \frac{\chi}{\omega + 1} + \lambda_1^3 & 
		\displaystyle\lambda_2^2 \frac{\chi}{\omega + 1} + \lambda_2^3 & 
		\displaystyle\lambda_3^2 \frac{\chi}{\omega + 1} + \lambda_3^3 & 
		\displaystyle\lambda_4^2 \frac{\chi}{\omega + 1} + \lambda_4^3 \\[3mm]
		\displaystyle\lambda_1 \ee^{\lambda_1} \frac{\chi}{\omega + 1} + \lambda_1^2 \ee^{\lambda_1} & 
		\displaystyle\lambda_2 \ee^{\lambda_2} \frac{\chi}{\omega + 1} + \lambda_2^2 \ee^{\lambda_2} & 
		\displaystyle\lambda_3 \ee^{\lambda_3} \frac{\chi}{\omega + 1} + \lambda_3^2 \ee^{\lambda_3} & 
		\displaystyle\lambda_4 \ee^{\lambda_4} \frac{\chi}{\omega + 1} + \lambda_4^2 \ee^{\lambda_4} \\[3mm]
		\displaystyle\lambda_1^2 \ee^{\lambda_1} \frac{\chi}{\omega + 1} + \lambda_1^3 \ee^{\lambda_1} & 
		\displaystyle\lambda_2^2 \ee^{\lambda_2} \frac{\chi}{\omega + 1} + \lambda_2^3 \ee^{\lambda_2} & 
		\displaystyle\lambda_3^2 \ee^{\lambda_3} \frac{\chi}{\omega + 1} + \lambda_3^3 \ee^{\lambda_3} & 
		\displaystyle\lambda_4^2 \ee^{\lambda_4} \frac{\chi}{\omega + 1} + \lambda_4^3 \ee^{\lambda_4}
	\end{bmatrix}
	\begin{bmatrix}
		c_1 \\[1mm]
		c_2 \\[1mm]
		c_3 \\[1mm]
		c_4 \\[1mm]
	\end{bmatrix}
	= 
	\begin{bmatrix}
		0 \\[1mm]
		0 \\[1mm]
		0 \\[1mm]
		0 \\[1mm]
	\end{bmatrix}.
\end{equation}
The linear system above admits non-trivial solutions -- corresponding to nontrivial eigenmodes -- only if the coefficient matrix is degenerate, that is, only if its determinant vanishes. 
This condition provides the characteristic equation for the circular frequency $\omega$ as a function of the dimensionless groups $\chi$ and $\eta_\perp$ 
\begin{equation}
	f(\omega\,;\chi,\eta_\perp) = 0,
\end{equation}
where 
\begin{equation}
	\label{eq:char_eq_f}
	\begin{aligned}
		f(\omega\,;\chi,\eta_\perp) &= \frac{1}{(\omega +1)^4} 
		\lambda _1 \lambda _2 \lambda _3 \lambda _4 
		\left[
			\lambda _1 (\omega +1)+\chi 
		\right] 
		\left[
			\lambda _2 (\omega +1)+\chi 
		\right] 
		\left[
			\lambda _3 (\omega +1)+\chi 
		\right] 
		\left[
			\lambda _4 (\omega +1)+\chi 
		\right] \times \\[2mm]
		&\times\left[
			\left(\ee^{\lambda_1} - \ee^{\lambda_2}\right) 
			\left(\ee^{\lambda_3} - \ee^{\lambda_4}\right) 
			(\lambda_1 \lambda_2 + \lambda_3 \lambda_4) -
			\left(\ee^{\lambda_1} - \ee^{\lambda_3}\right) 
			\left(\ee^{\lambda_2} - \ee^{\lambda_4}\right) 
			(\lambda_1 \lambda_3 + \lambda_2 \lambda_4) \,+
		\right. \\[2mm]
		&\left.
			+
            \left(\ee^{\lambda_1} - \ee^{\lambda_4}\right)
			\left(\ee^{\lambda_2} - \ee^{\lambda_3}\right) 
			(\lambda_1 \lambda_4 + \lambda_2 \lambda_3)
		\right].
	\end{aligned}
\end{equation}

\subsection{Computation of the stability map}
\label{LIN:ALG}
The distribution of the roots of the characteristic equation \eqref{LIN:eq:char_eq} in the complex plane was graphically explored for a fixed choice of the parameters $\chi$ and $\eta_\perp$ corresponding to flutter instability in numerical simulations.
Having thus retrieved an estimate for the location of the leading eigenvalue $\omega_{\max}$ for such a choice of model parameters, this was used as initial guess for the Newton-like solver \textsf{FindRoot} of \textsf{Wolfram Mathematica~13.2}.

To explore the stability of the system in the plane of parameters $(\eta_\perp,\chi) \in [0,350]\times[-130,130]$ and construct the stability map shown in \Fig\ref{fig:stability}\ref{fig:stability:stability_map}, we started from the solution determined above and exploited a root-following algorithm while varying the parameters:
\smallskip
\textcolor{gray}{
\begin{algorithmic}[1]
	\State compute $\omega_{\max}$ for fixed parameters $(\bar{\eta}_\perp,\bar{\chi})$
	\State choose a step size $\Delta_{\eta_\perp}$ and a number of steps $N_{\eta_\perp}$ for the discretization of $\eta_\perp$
	\State choose a step size $\Delta_{\chi}$ and a number of steps $N_{\chi}$ for the discretization of $\chi$
	\For{$i = 1$ to $N_{\eta_\perp}$}
		\State $\eta_\perp \gets \bar{\eta}_\perp + i \Delta_{\eta_\perp}$ \label{alg:following:eta_upd}
		\For{$j = 1$ to $N_{\chi}$}
			\State $\chi \gets \bar{\chi} + j \Delta_{\chi}$ \label{alg:following:chi_upd}
			\State compute $\omega_{\max}'$ for the new parameters $(\eta_\perp,\chi)$ using $\omega_{\max}$ as initial guess
			\State $\omega_{\max} \gets \omega_{\max}'$
			\State save the value of $\omega_{\max}$
		\EndFor
	\EndFor
\end{algorithmic}
}
\smallskip
The above algorithm takes care of following the leading eigenvalue in the quadrant $\eta_\perp > \bar{\eta}_\perp$ and $\chi > \bar{\chi}$, but suitably changing the sign in parameters updates at lines~\ref{alg:following:eta_upd} and \ref{alg:following:chi_upd} allows to follow the eigenvalue in all directions. 

We remark that to ensure convergence of the Newton iterations to the right root, \ie, the one corresponding to the same eigenfunction, the steps sizes $\Delta_{\eta_\perp}$ and $\Delta_{\chi}$ need to be sufficiently small. The absence of discontinuities in the curves reported in \Fig\ref{fig:stability}\ref{fig:stability:bifurcation_plot} indicates that our choice of $\Delta_{\eta_\perp} = \Delta_{\chi} = \num{1e-1}$ was fine enough to avoid jumps in the root-following algorithm from one branch of solutions to another.

The eigenmodes shown in \Figs\ref{fig:stability}\ref{fig:stability:mode_1}--\ref{fig:stability:mode_2} are determined from \eqref{LIN:eq:lin_ODE_sol} with the vector of integration constants $\{c_1,c_2,c_3,c_4\}$ in the kernel of the matrix in \eqref{LIN:eq:constr_matrix}. 

% Bibliography
\bibliographystyle{elsarticle-num}
\bibliography{references.bib}

\begin{thebibliography}{10}
\expandafter\ifx\csname url\endcsname\relax
  \def\url#1{\texttt{#1}}\fi
\expandafter\ifx\csname urlprefix\endcsname\relax\def\urlprefix{URL }\fi
\expandafter\ifx\csname href\endcsname\relax
  \def\href#1#2{#2} \def\path#1{#1}\fi

\bibitem{oiwa_2003}
S.~A. Burgess, M.~L. Walker, H.~Sakakibara, P.~J. Knight, K.~Oiwa, Dynein structure and power stroke, Nature 421~(6924) (2003) 715--718.
\newblock \href {https://doi.org/10.1038/nature01377} {\path{doi:10.1038/nature01377}}.

\bibitem{nicastro_2014}
J.~Lin, K.~Okada, M.~Raytchev, M.~C. Smith, D.~Nicastro, Structural mechanism of the dynein power stroke, Nature Cell Biology 16~(5) (2014) 479--485.
\newblock \href {https://doi.org/10.1038/ncb2939} {\path{doi:10.1038/ncb2939}}.

\bibitem{nicastro_2018}
J.~Lin, D.~Nicastro, Asymmetric distribution and spatial switching of dynein activity generates ciliary motility, Science 360~(6387) (2018) eaar1968.
\newblock \href {https://doi.org/10.1126/science.aar1968} {\path{doi:10.1126/science.aar1968}}.

\bibitem{purcell_1977}
E.~M. Purcell, Life at low {R}eynolds number, American Journal of Physics 45~(1) (1977) 3--11.
\newblock \href {https://doi.org/10.1119/1.10903} {\path{doi:10.1119/1.10903}}.

\bibitem{Lauga2011_Life}
E.~Lauga, Life around the scallop theorem, Soft Matter 7 (2011) 3060--3065.
\newblock \href {https://doi.org/10.1039/C0SM00953A} {\path{doi:10.1039/C0SM00953A}}.

\bibitem{Yuan2021}
S.~Yuan, Z.~Wang, H.~Peng, S.~M. Ward, G.~W. Hennig, H.~Zheng, W.~Yan, Oviductal motile cilia are essential for oocyte pickup but dispensable for sperm and embryo transport, Proceedings of the National Academy of Sciences 118~(22) (2021) e2102940118.
\newblock \href {https://doi.org/10.1073/pnas.2102940118} {\path{doi:10.1073/pnas.2102940118}}.

\bibitem{Dreyfus2005_Microscopic}
R.~Dreyfus, J.~Baudry, M.~L. Roper, M.~Fermigier, H.~A. Stone, J.~Bibette, Microscopic artificial swimmers, Nature 437~(7060) (2005) 862--865.
\newblock \href {https://doi.org/10.1038/nature04090} {\path{doi:10.1038/nature04090}}.

\bibitem{Lum2016_Shape}
G.~Z. Lum, Z.~Ye, X.~Dong, H.~Marvi, O.~Erin, W.~Hu, M.~Sitti, Shape-programmable magnetic soft matter, Proceedings of the National Academy of Sciences 113~(41) (2016) E6007--E6015.
\newblock \href {https://doi.org/10.1073/pnas.1608193113} {\path{doi:10.1073/pnas.1608193113}}.

\bibitem{Kaynak2017_Acoustic}
M.~Kaynak, A.~Ozcelik, A.~Nourhani, P.~E. Lammert, V.~H. Crespi, T.~J. Huang, Acoustic actuation of bioinspired microswimmers, Lab on a Chip 17~(3) (2017) 395--400.
\newblock \href {https://doi.org/10.1039/c6lc01272h} {\path{doi:10.1039/c6lc01272h}}.

\bibitem{Toonder2008_Artificial}
J.~d. Toonder, F.~Bos, D.~Broer, L.~Filippini, M.~Gillies, J.~de~Goede, T.~Mol, M.~Reijme, W.~Talen, H.~Wilderbeek, V.~Khatavkar, P.~Anderson, Artificial cilia for active micro-fluidic mixing, Lab on a Chip 8~(4) (2008) 533.
\newblock \href {https://doi.org/10.1039/b717681c} {\path{doi:10.1039/b717681c}}.

\bibitem{Shields2010_Biomimetic}
A.~R. Shields, B.~L. Fiser, B.~A. Evans, M.~R. Falvo, S.~Washburn, R.~Superfine, Biomimetic cilia arrays generate simultaneous pumping and mixing regimes, Proceedings of the National Academy of Sciences 107~(36) (2010) 15670--15675.
\newblock \href {https://doi.org/10.1073/pnas.1005127107} {\path{doi:10.1073/pnas.1005127107}}.

\bibitem{Vilfan2010_Self}
M.~Vilfan, A.~Potočnik, B.~Kavčič, N.~Osterman, I.~Poberaj, A.~Vilfan, D.~Babič, Self-assembled artificial cilia, Proceedings of the National Academy of Sciences 107~(5) (2010) 1844--1847.
\newblock \href {https://doi.org/10.1073/pnas.0906819106} {\path{doi:10.1073/pnas.0906819106}}.

\bibitem{Dong2020_Bioinspired}
X.~Dong, G.~Z. Lum, W.~Hu, R.~Zhang, Z.~Ren, P.~R. Onck, M.~Sitti, Bioinspired cilia arrays with programmable nonreciprocal motion and metachronal coordination, Science Advances 6~(45) (Nov. 2020).
\newblock \href {https://doi.org/10.1126/sciadv.abc9323} {\path{doi:10.1126/sciadv.abc9323}}.

\bibitem{lesich_2010}
C.~B. Lindemann, K.~A. Lesich, Flagellar and ciliary beating: the proven and the possible, Journal of Cell Science 123~(4) (2010) 519--528.
\newblock \href {https://doi.org/10.1242/jcs.051326} {\path{doi:10.1242/jcs.051326}}.

\bibitem{bayly_2015}
P.~V. Bayly, K.~S. Wilson, Analysis of unstable modes distinguishes mathematical models of flagellar motion, Journal of The Royal Society Interface 12~(106) (2015) 20150124.
\newblock \href {https://doi.org/10.1098/rsif.2015.0124} {\path{doi:10.1098/rsif.2015.0124}}.

\bibitem{howard_2016}
P.~Sartori, V.~F. Geyer, A.~Scholich, F.~Jülicher, J.~Howard, Dynamic curvature regulation accounts for the symmetric and asymmetric beats of \textit{Chlamydomonas} flagella, eLife 5 (2016) e13258.
\newblock \href {https://doi.org/10.7554/eLife.13258} {\path{doi:10.7554/eLife.13258}}.

\bibitem{Maeda2007_SelfWalking}
S.~Maeda, Y.~Hara, T.~Sakai, R.~Yoshida, S.~Hashimoto, Self‐walking gel, Advanced Materials 19~(21) (2007) 3480--3484.
\newblock \href {https://doi.org/10.1002/adma.200700625} {\path{doi:10.1002/adma.200700625}}.

\bibitem{zhao_2019}
Y.~Zhao, C.~Xuan, X.~Qian, Y.~Alsaid, M.~Hua, L.~Jin, X.~He, Soft phototactic swimmer based on self-sustained hydrogel oscillator, Science Robotics 4~(33) (2019) eaax7112.
\newblock \href {https://doi.org/10.1126/scirobotics.aax7112} {\path{doi:10.1126/scirobotics.aax7112}}.

\bibitem{korner_2020}
K.~Korner, A.~S. Kuenstler, R.~C. Hayward, B.~Audoly, K.~Bhattacharya, A nonlinear beam model of photomotile structures, Proceedings of the National Academy of Sciences 117~(18) (2020) 9762--9770.
\newblock \href {https://doi.org/10.1073/pnas.1915374117} {\path{doi:10.1073/pnas.1915374117}}.

\bibitem{stone_2020}
L.~Zhu, H.~A. Stone, Harnessing elasticity to generate self-oscillation via an electrohydrodynamic instability, Journal of Fluid Mechanics 888 (2020) A31.
\newblock \href {https://doi.org/10.1017/jfm.2020.54} {\path{doi:10.1017/jfm.2020.54}}.

\bibitem{stone_2021}
E.~Han, L.~Zhu, J.~W. Shaevitz, H.~A. Stone, Low-{R}eynolds-number, biflagellated {Q}uincke swimmers with multiple forms of motion, Proceedings of the National Academy of Sciences 118~(29) (2021) e2022000118.
\newblock \href {https://doi.org/10.1073/pnas.2022000118} {\path{doi:10.1073/pnas.2022000118}}.

\bibitem{Doi1992}
M.~Doi, M.~Matsumoto, Y.~Hirose, Deformation of ionic polymer gels by electric fields, Macromolecules 25~(20) (1992) 5504--5511.
\newblock \href {https://doi.org/10.1021/ma00046a058} {\path{doi:10.1021/ma00046a058}}.

\bibitem{Hong2010}
W.~Hong, X.~Zhao, Z.~Suo, Large deformation and electrochemistry of polyelectrolyte gels, Journal of the Mechanics and Physics of Solids 58~(4) (2010) 558--577.
\newblock \href {https://doi.org/10.1016/j.jmps.2010.01.005} {\path{doi:10.1016/j.jmps.2010.01.005}}.

\bibitem{Gray1955}
J.~Gray, G.~J. Hancock, The propulsion of sea-urchin spermatozoa, Journal of Experimental Biology 32~(4) (1955) 802--814.
\newblock \href {https://doi.org/10.1242/jeb.32.4.802} {\path{doi:10.1242/jeb.32.4.802}}.

\bibitem{Lighthill1976}
J.~Lighthill, Flagellar hydrodynamics, {SIAM} Review 18~(2) (1976) 161--230.
\newblock \href {https://doi.org/10.1137/1018040} {\path{doi:10.1137/1018040}}.

\bibitem{koens_2017}
L.~Koens, E.~Lauga, Analytical solutions to slender-ribbon theory, Physical Review Fluids 2~(8) (2017) 084101.
\newblock \href {https://doi.org/10.1103/PhysRevFluids.2.084101} {\path{doi:10.1103/PhysRevFluids.2.084101}}.

\bibitem{Goriely2017}
A.~Goriely, The Mathematics and Mechanics of Biological Growth, Springer, New York, NY, 2017.
\newblock \href {https://doi.org/10.1007/978-0-387-87710-5} {\path{doi:10.1007/978-0-387-87710-5}}.

\bibitem{Ziegler1977}
H.~Ziegler, Principles of structural stability, Springer Basel AG, 1977.
\newblock \href {https://doi.org/10.1007/978-3-0348-5912-7} {\path{doi:10.1007/978-3-0348-5912-7}}.

\bibitem{bigoni2023flutter}
D.~Bigoni, F.~Dal~Corso, O.~N. Kirillov, D.~Misseroni, G.~Noselli, A.~Piccolroaz, Flutter instability in solids and structures, with a view on biomechanics and metamaterials, Proceedings of the Royal Society A 479~(2279) (2023) 20230523.
\newblock \href {https://doi.org/10.1098/rspa.2023.0523} {\path{doi:10.1098/rspa.2023.0523}}.

\bibitem{Damioli2022}
V.~Damioli, E.~Zorzin, A.~DeSimone, G.~Noselli, A.~Lucantonio, Transient shape morphing of active gel plates: geometry and physics, Soft Matter 18~(31) (2022) 5867--5876.
\newblock \href {https://doi.org/10.1039/d2sm00669c} {\path{doi:10.1039/d2sm00669c}}.

\bibitem{Strogatz2019}
S.~Strogatz, Nonlinear dynamics and chaos, CRC Press, Boca Raton, 2019.
\newblock \href {https://doi.org/10.1201/9780429492563} {\path{doi:10.1201/9780429492563}}.

\bibitem{Camalet2000}
S.~Camalet, F.~Jülicher, Generic aspects of axonemal beating, New Journal of Physics 2~(1) (2000) 324.
\newblock \href {https://doi.org/10.1088/1367-2630/2/1/324} {\path{doi:10.1088/1367-2630/2/1/324}}.

\bibitem{antman2005}
S.~S. Antman, Nonlinear problems of elasticity, Springer, New York, NY, 2005.
\newblock \href {https://doi.org/10.1007/0-387-27649-1} {\path{doi:10.1007/0-387-27649-1}}.

\bibitem{cicconofri_2023}
G.~Cicconofri, V.~Damioli, G.~Noselli, Nonreciprocal oscillations of polyelectrolyte gel filaments subject to a steady and uniform electric field, Journal of the Mechanics and Physics of Solids 173 (2023) 105225.
\newblock \href {https://doi.org/10.1016/j.jmps.2023.105225} {\path{doi:10.1016/j.jmps.2023.105225}}.

\bibitem{smith_2010}
M.~B. Smith, H.~Li, T.~Shen, X.~Huang, E.~Yusuf, D.~Vavylonis, Segmentation and tracking of cytoskeletal filaments using open active contours, Cytoskeleton 67~(11) (2010) 693--705.
\newblock \href {https://doi.org/10.1002/cm.20481} {\path{doi:10.1002/cm.20481}}.

\end{thebibliography}

\pagebreak
\section*{Supplementary videos}

\noindent The supplementary videos are available at:\,\url{https://hdl.handle.net/20.500.12928/SD0KIF}.

\bigskip

\begin{movie}[Swimming PEH ribbon in constant electric field from flutter instability.]	
    \label{MOV:exp_swimming}
    \begin{minipage}[t]{0.55\linewidth}
        \vspace{0pt}
        Swimming of a PEH ribbon under a steady and uniform electric field. In the experiment, the applied electric potential is $\Delta V = \qty{85}{\V}$. The self-sustained, periodic shape changes are nonreciprocal and resemble a backward traveling wave of increasing amplitude, as highlighted by the kymograph.
    \end{minipage}
    \hfill
    \begin{minipage}[t]{0.4\linewidth}
        \vspace{0pt}
        \raggedleft
        \includegraphics[width=\linewidth]{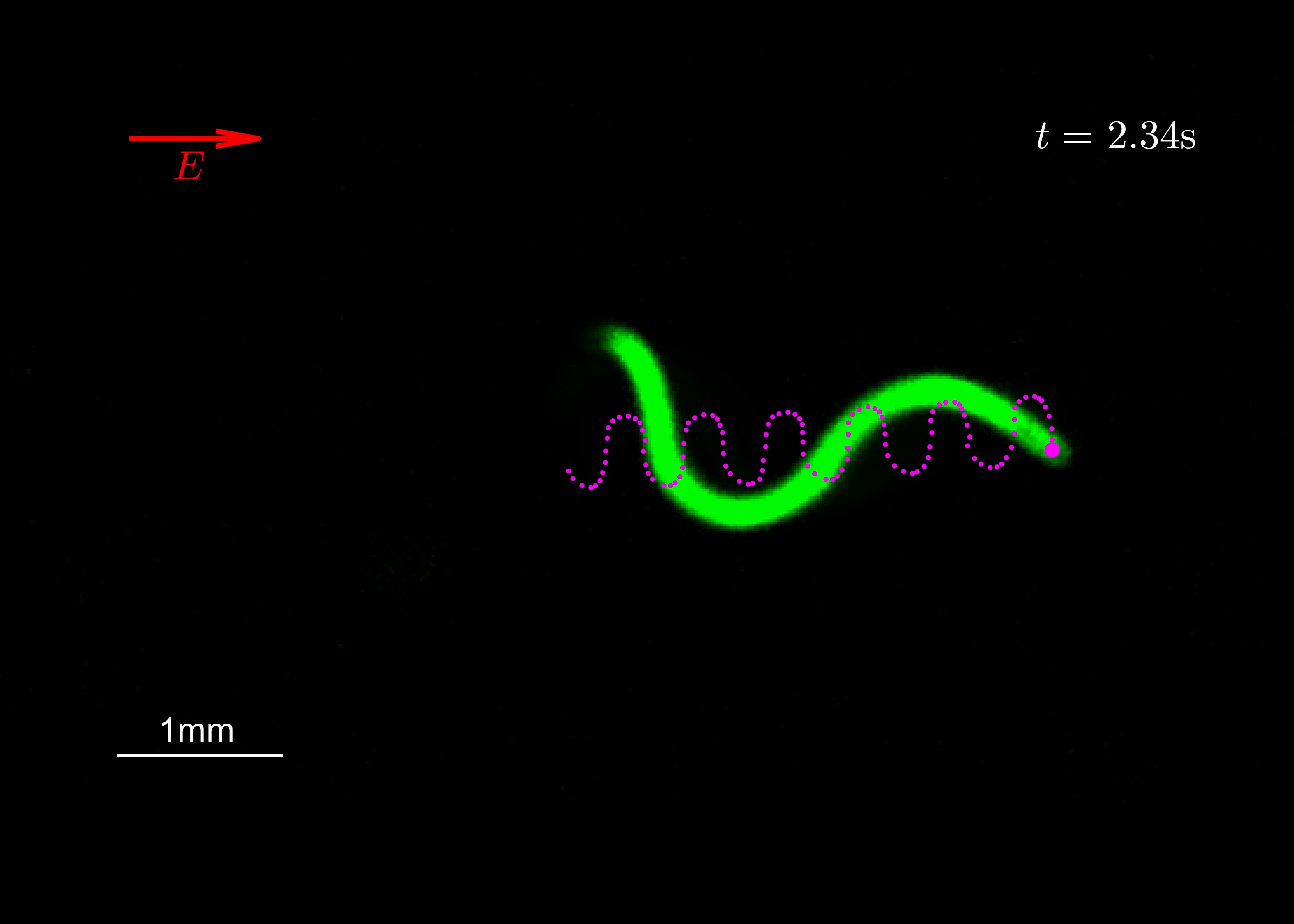}
    \end{minipage}
\end{movie}

\begin{movie}[Activity of PEH ribbons in constant electric field.]
    \label{MOV:exp_stationary}
    \begin{minipage}[t]{0.55\linewidth}
        \vspace{0pt}
        Active reconfiguration of PEH filaments under subcritical electric potential $\Delta V = \qty{40}{V}$ for different initial inclinations to the electric field (red arrow). 
        The curvature of the filament develops by bending towards the cathode in a characteristic time -- which is common to all filaments -- and its distribution at equilibrium depends on the local filament's orientation in the electric field.
        The maximum curvature is attained where the field lines cross the filament perpendicularly and is the same for all initial conditions, excluding the one aligned with the field, for which the filament does not develop curvature. 
    \end{minipage}
    \hfill
    \begin{minipage}[t]{0.4\linewidth}
        \vspace{0pt}
        \raggedleft
        \includegraphics[width=\linewidth]{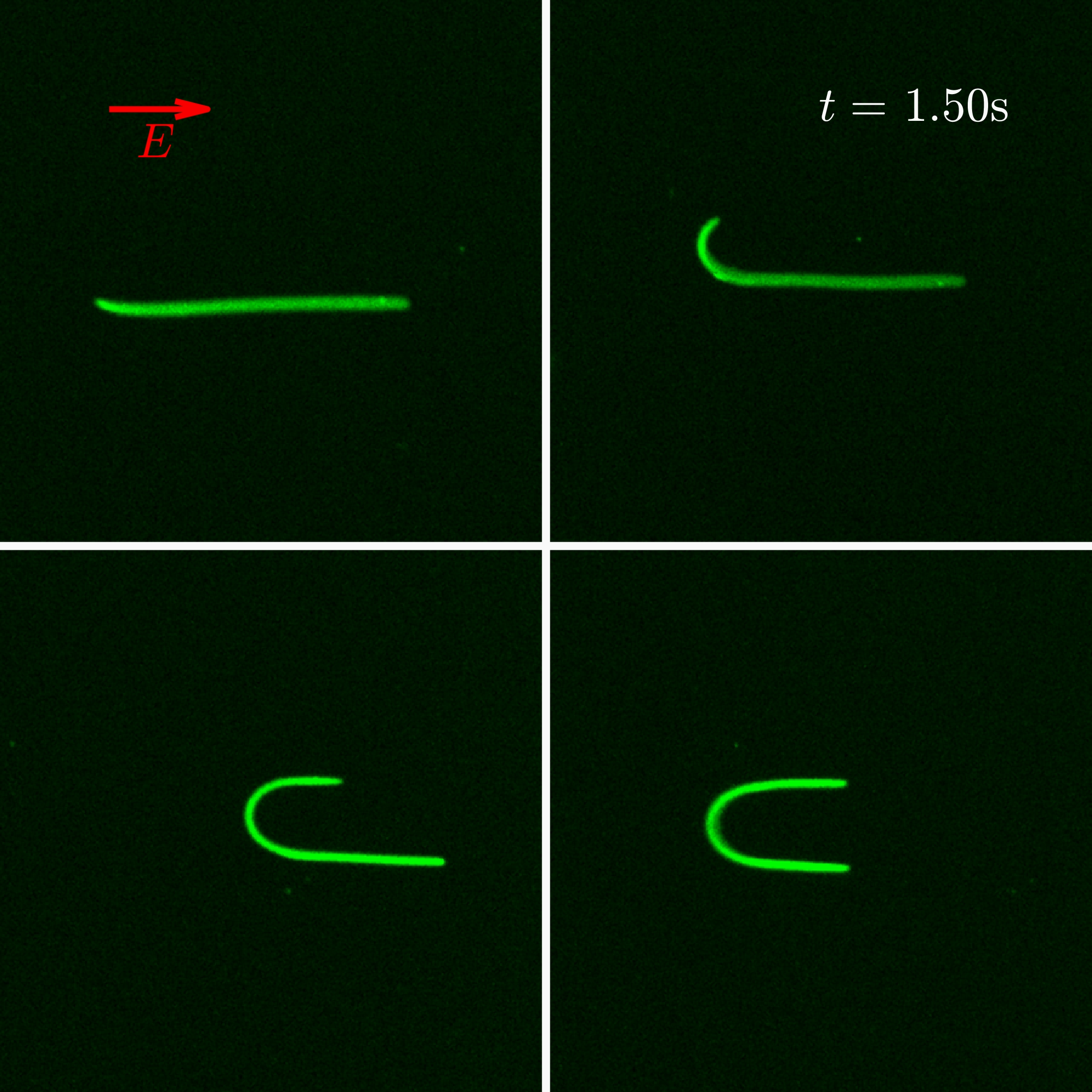}
    \end{minipage}
\end{movie}

\begin{movie}[Onset of flutter instability in PEH ribbon under constant electric field.]
    \label{MOV:exp_onset}
    \begin{minipage}[t]{0.55\linewidth}
        \vspace{0pt}
        Motion of a PEH ribbon at the onset of flutter instability. In the experiment, the applied electric potential is $\Delta V = \qty{85}{\V}$.
        Starting from a  straight equilibrium configuration, activation of the electric field results in oscillations of increasing amplitude. Initially, small amplitude oscillations generate no thrust, but they do so once the motion has developed into large amplitude periodic oscillations, as revealed by the kymographs.
    \end{minipage}
    \hfill
    \begin{minipage}[t]{0.4\linewidth}
        \vspace{0pt}
        \raggedleft
        \includegraphics[width=\linewidth]{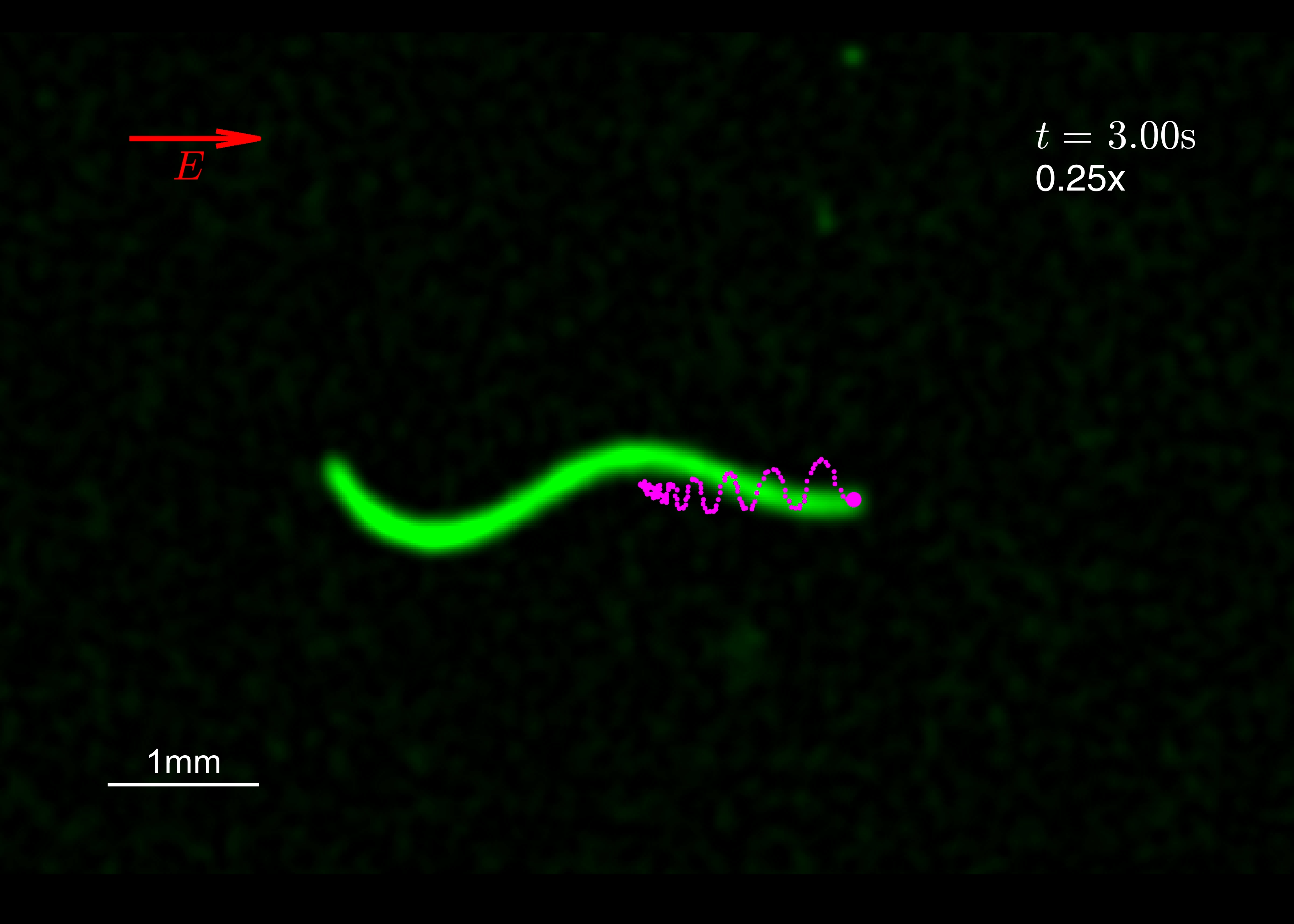}
    \end{minipage}
\end{movie}

\pagebreak

\begin{movie}[Swimming of morphoelastic filament originating from flutter instability.]
    \label{MOV:swimmer_virtual}
    \begin{minipage}[t]{0.55\linewidth}
        \vspace{0pt}
        Virtual swimmer from instability onset to steady oscillations and related kymographs.
        Numerical simulations with parameters extracted from the representative experiment shown in \mov\ref{MOV:exp_stationary} as detailed in \sect\ref{NUM:FIT}.
        Starting from a straight equilibrium configuration, upon activation of the external stimulus the virtual filament exhibits oscillations, which are initially of small amplitude and generate no propulsion. Such oscillations rapidly increase in amplitude to attain a limit cycle. Such periodic shape change allows the filament to swim. \\
        The kymographs at the onset of the instability show the emergence of small amplitude oscillations close to the linear regime. 
        The kymographs during the periodic motion highlight the similarity between the shape change of the virtual swimmer and that observed in the physical experiment of \mov\ref{MOV:exp_stationary}.
    \end{minipage}
    \hfill
    \begin{minipage}[t]{0.4\linewidth}
        \vspace{0pt}
        \raggedleft
        \includegraphics[width=\linewidth]{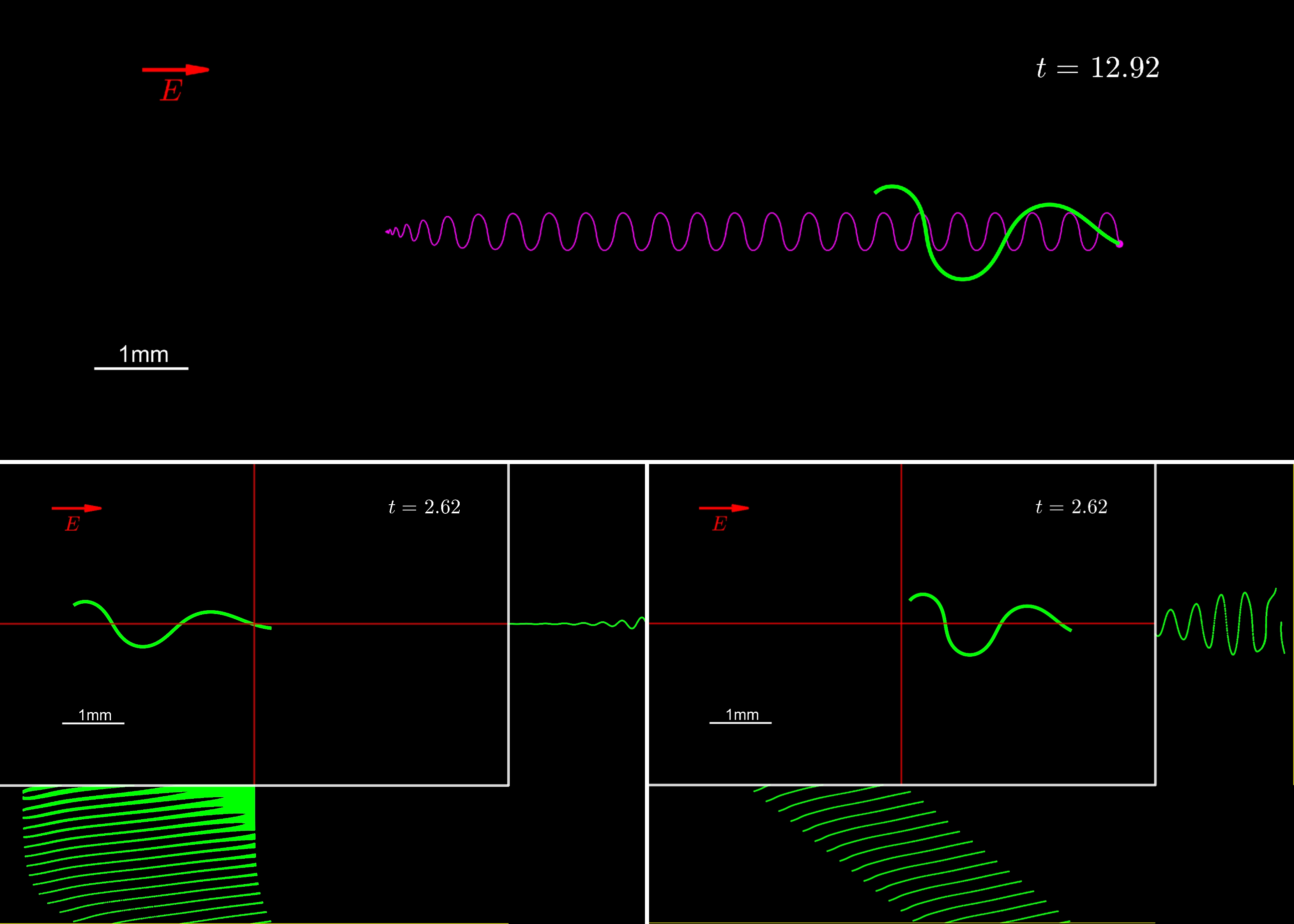}
    \end{minipage}
\end{movie}

\begin{movie}[Steering the virtual swimmer in the fluid domain.]
    \label{MOV:circus}
    \begin{minipage}[t]{0.55\linewidth}
        \vspace{0pt}
        The virtual swimmer can be steered in the fluid domain by simply reorienting the external stimulus, as described in \sect\ref{MOD}. This video reports the prototypical case in which the swimmer is directed along a circular trajectory by changing the stimulus direction at constant angular velocity, as reported in the lover panel of \Fig\ref{fig:trajectories}. All the model parameters are as in \tab\ref{MOD:SIM:tab:parameters}.
    \end{minipage}
    \hfill
    \begin{minipage}[t]{0.4\linewidth}
        \vspace{0pt}
        \includegraphics[width=\linewidth]{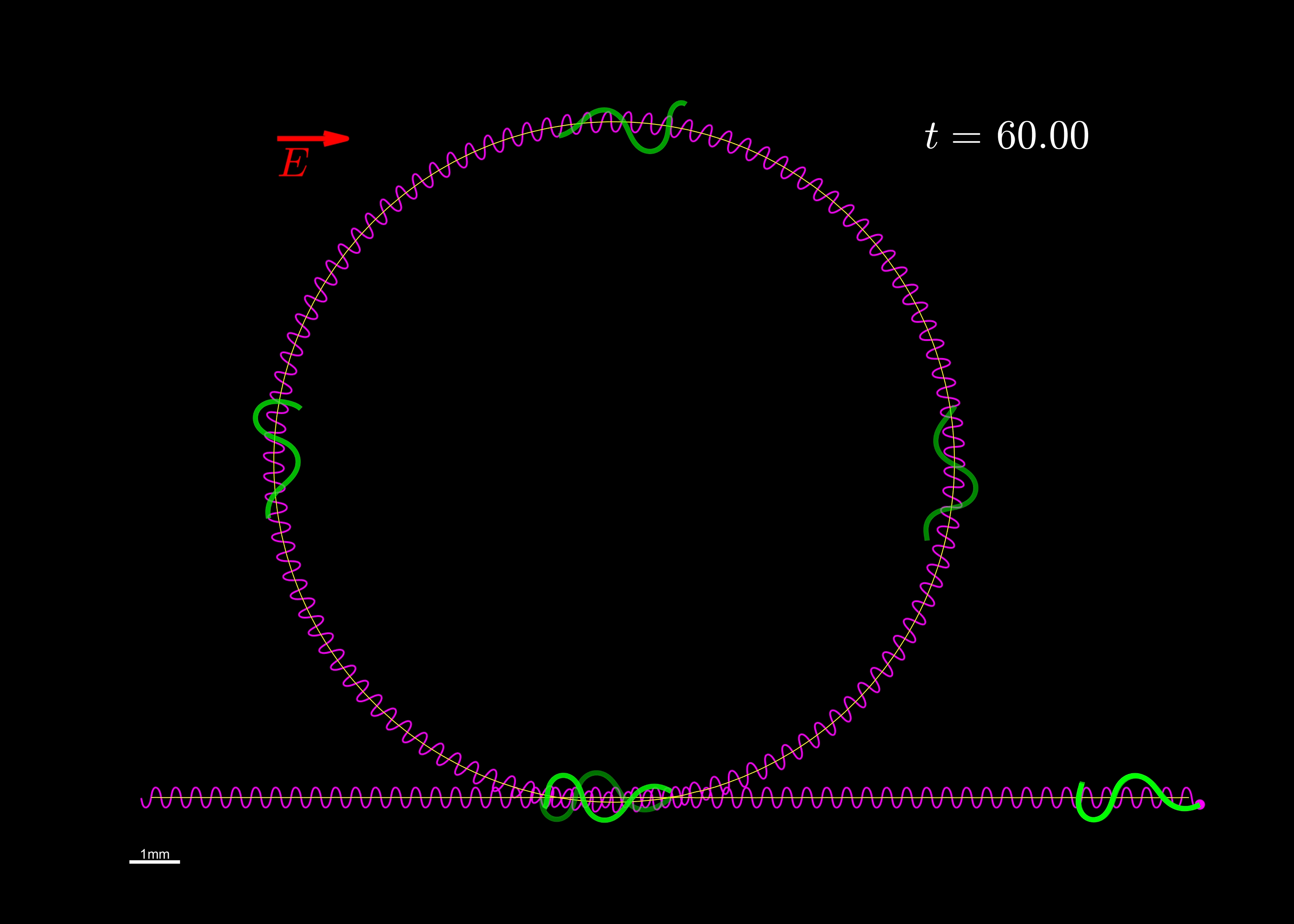}
    \end{minipage}
\end{movie}

\end{document}